\documentclass[prd,aps,nofootinbib,floatfix,10pt]{revtex4}
\usepackage{amsmath,graphicx,epsfig,amssymb,dsfont,mathtools}
\usepackage[usenames]{color}

\def\mI{m_{\Lambda_{b}}}
\def\mf{m_{\Lambda^{*}_J}}
\def\mjp{m_{J/\psi}}

\begin{document}
\title{Angular distributions for  $\Lambda_b\to \Lambda^*_J (pK^-)J/\psi(\to \ell^+\ell^-)$ Decays}
\author{
Zhi-Peng Xing$^{1}$~\footnote{Email:zpxing@sjtu.edu.cn}, 
Fei Huang$^{2}$~\footnote{Email:fhuang@sjtu.edu.cn},
Wei Wang$^{2}$~\footnote{Email:wei.wang@sjtu.edu.cn}}

\affiliation{$^{1}$ Tsung-Dao Lee Institute, Shanghai Jiao Tong University, Shanghai 200240, China}
\affiliation{$^{2}$ INPAC, Shanghai Key Laboratory for Particle Physics and Cosmology,\\
Key Laboratory for Particle Astrophysics and Cosmology (MOE),\\
School of Physics and Astronomy, Shanghai Jiao-Tong University, Shanghai
200240, P.R. China }

\begin{abstract}
We carry out an analysis  of the multibody decay cascade   $\Lambda_b\to  \Lambda^*_J J/\psi \to  pK^-  J/\psi$  with the $  \Lambda^*_J$ resonance including  $\Lambda^*_{1520},\Lambda^*_{1600}$, and $\Lambda^*_{1800}$,  and $J/\psi$ reconstructed by the lepton pair final state.  Using the helicity amplitude technique, we derive a compact form for the angular distributions for the decay chain, from which one can extract various one-dimensional distributions.  Using the $\Lambda_b\to \Lambda_J^*$ form factors from lattice QCD and quark model, we calculate the differential and integrated partial widths. Decay branching fractions  are found as $\mathcal{B}(\Lambda_b\to  \Lambda^*_J( pK)J/\psi(\ell^+\ell^-))=(1.35\pm0.28)\times 10^{-5}$.  In addition, we also explore forward-backward asymmetry, and various polarizations. Results in this work will  serve a calibration for the study of  $b\to s  \ell^+\ell^-$ decays in $\Lambda_b$ decays in future and provide useful information towards the understanding of the properties of the $\Lambda_J^*$ baryons. 
\end{abstract}
\maketitle

\section{Introduction}

Multi-body hadronic decays of heavy mesons and baryons are of special interest due to various reasons. Compared to two-body hadronic decay, multi-body decays typically  have much richer phase spaces, and thus can be used to explore various new phenomena. Since these decays might receive distinct resonating  contributions, they   provide a platform for the study of strong interactions and the examination of the beneath quantum field theory, i.e. quantum chromodynamics (QCD),  in a versatile manner. In addition, in the past decades, many traditional and exotic hadron structures are discovered in multi-body decays of heavy mesons and baryons at different experimental facilities~\cite{Belle:2003nnu,Belle:2004lle,BaBar:2004oro,LHCb:2015yax,LHCb:2020jpq}.


The main focus of this work is the  $\Lambda_b \to pK^-J/\psi$ decay, which has been previously  explored on the experimental side. This process plays a very important role in the search for exotic hadron states.
In 2015, the LHCb collaboration has reported two exotic structures,  $P_c(4380)$ and $P_{c}(4450)$,  firstly observed in the $\Lambda_{b}^{0}\to J/\psi p K^{-}$ process~\cite{LHCb:2015yax}. In addition, a new narrow state $P_c(4312)$ and a two-peak structure of $P_c(4450)$ have been discovered by analyzing the  $\Lambda_{b}^{0}\to J/\psi p K^{-}$ data from the LHCb collaboration~\cite{LHCb:2019kea}.  While the $P_c$ resonances give a sizable contributions to the decay widths, the $\Lambda_b\to \Lambda^*_J (pK^-)J/\psi$ contributions are also likely significant. Thus the identification of exotic hadrons and precise determinations of their properties strongly depend on the understanding of the dynamics in this decay process. Actually, the contribution from the $P_{c}$ pentaquark is    small  in the low-invariant mass range ($M_{pk}$=$1.4{\rm GeV}\sim1.8{\rm GeV}$)  while the  $\Lambda_{1520, 1600, 1800}^{*}$ resonances  occupy dominant contributions in this energy range~\cite{LHCb:2015yax}.  In this work,  we mainly focus on  the  $\Lambda_{J}^{*}$ resonance contributions.

Another salient feature of the  $\Lambda_{b}\rightarrow{\Lambda^*_J({p K})~J/\psi({\ell^{+}\ell^{-}})}$ decay is the wealth  information carried by angular  observables in terms of angular asymmetries that can be used to probe new physics (NP) beyond the standard model (BSM). Our process is the basis for Flavour-changing neutral current(FCNC) processes which is involved in the wilson coefficient $C_9^{eff}$ in Ref.~\cite{Buras:1994dj}.
The FCNC process of $b\rightarrow{s \ell^{+}\ell^{-}}$ is forbidden at the tree level and thus sensitive to new physics beyond the standard model. Thus the $B\to K\ell^+\ell^-$  and  $B\to K^*\ell^+\ell^-$ have received great attention in the past decades and have provided very stringent constraints on new physics beyond the standard model~\cite{Belle:2009zue,He:2009tf,Xing:2018lre,Zhao:2018mrg,Huber:2019iqf,Huber:2020vup,MunirBhutta:2020ber,Li:2021qyo,He:2021yoz,Cen:2021iwv,Li:2022nim}. 
Meanwhile in these decays, the so-called flavor anomalies are also found~\cite{BaBar:2012obs,Belle:2016fev,LHCb:2017avl,LHCb:2015gmp,LHCb:2018jna,LHCb:2019efc,LHCb:2021xxq,LHCb:2021trn,LHCb:2021awg}. For instance,  LHCb has presented its latest measurement of the ratio of branching fractions~\cite{LHCb:2021trn,LHCb:2017avl}:
\begin{eqnarray}
R_K &\equiv& \frac{{\cal B}(B\to K\mu^+\mu^-)}{{\cal B}(B\to Ke^+e^-)}=0.846^{+0.042+0.013}_{-0.039-0.012},\qquad 1.1\rm{GeV}^2<q^2<6\rm{GeV}^2,\nonumber\\
R_{K^*}&\equiv& \frac{{\cal B}(B\to K^*\mu^+\mu^-)}{{\cal B}(B\to K^*e^+e^-)}=0.69^{+0.11}_{-0.07}\pm 0.05,\qquad 1.1\rm{GeV}^2<q^2<6\rm{GeV}^2,
\end{eqnarray}
which has the $3.1\sigma$ and $2.8\sigma$ tension with the  SM prediction, respectively.  To further examine the implication of these observations, more experimental and theoretical analyses are called for. The $\Lambda_b$ is a spin-1/2 hadron and has more polarization degrees of freedom than $B$ meson and thus it is presumable that  the baryonic decay $\Lambda_b\to \Lambda^*_J(\to pK^-)\ell^+\ell^-$ provides complementary information. In this regard, a detailed analysis of $\Lambda_b\to \Lambda^*_J J/\psi(\to \ell^+\ell^-)$ can provide a valuable benchmark.

The focus of this paper is  the angular distributions for $\Lambda_{b}\to\Lambda^{*}_{J}(\to pK^{-})J/\psi(\to \ell^{+}\ell^{-})$, where $\Lambda^{*}_{J}$   can decay into  the $pK^{-}$ final state. The angular distributions for $\Lambda_b$ four-body decay with resonances depend on  different spin-parity of the $\Lambda^{*}_{J}$ resonance and the interference between them. Based on the relevant experimental data~\cite{LHCb:2015yax}, we find the resonances $\Lambda_{1405}^{*}$, $\Lambda_{1520}^{*}$, $\Lambda_{1600}^{*}$, $\Lambda_{1800}^{*}$ and $\Lambda_{1810}^{*}$ give main contributions  compared to other resonances, especially for $\Lambda^{*}_{1670}$ with tiny contributions and $\Lambda^{*}_{1690}$ with small integrated width.  Since  the  $\Lambda^{*}_{J}$ will  decay into p K, the  $\Lambda_J^*$ mass should be above  $m_K+m_p$ and thereby resonances like the  $\Lambda_{1405}^{*}$ are not allowed.  In addition, the $\Lambda^{*}_{1810}$ is very close to $\Lambda^{*}_{1800}$, and will be treated together in the following. Therefore we only consider three resonances $\Lambda_{1520}^{*}$, $\Lambda_{1600}^{*}$ and  $\Lambda_{1800}^{*}$ in our work. The spin-parity quantum numbers, masses, and decay widths of these resonances are shown in Table.~\ref{res}. 
 
\begin{table}[htbp!]
\caption{The spin-party, masses and decay width of resonance $\Lambda^*_{1520},\Lambda^*_{1600}$ and $\Lambda^*_{1800}$~\cite{Workman:2022}. }\label{res}
\begin{tabular}{|c|c|c|c|c|c|c|c|}\hline\hline
Resonance&$J^P$& Mass(${\rm MeV}$)&$\Gamma$(${\rm MeV}$)\\\hline \hline
$\Lambda^*_{1520}$&$\frac{3}{2}^-$&$1519.42\pm0.19$&$15.73\pm0.26$ \\\hline
$\Lambda^*_{1600}$&$\frac{1}{2}^+$&$\sim1600$&$\sim 200$ \\\hline
$\Lambda^*_{1800}$&$\frac{1}{2}^-$&$\sim1800$&$\sim200$ \\\hline
\end{tabular}
\end{table}


The rest of this paper is organized as follows. In Sec.II, we give the theoretical framework for the $\Lambda_{b}\to \Lambda_J^* J/\psi$ with the $\Lambda^{*}_J$ having different quantum numbers. The helicity amplitude is adopted to derive the angular distributions. In Sec.III, we make use of  $\Lambda_{b}\to \Lambda_J^*$ from Lattice QCD calculation and a quark model and calculate the differential decay widths. Angular distribution variables are also explored in this section, and in particular the forward-backward asymmetry and polarizations are predicted. A brief summary will be presented in the last section. Some calculation details are collected in the appendix.

\section{Helicity Amplitudes}

\begin{figure}[!h]
\includegraphics[width=0.7\columnwidth]{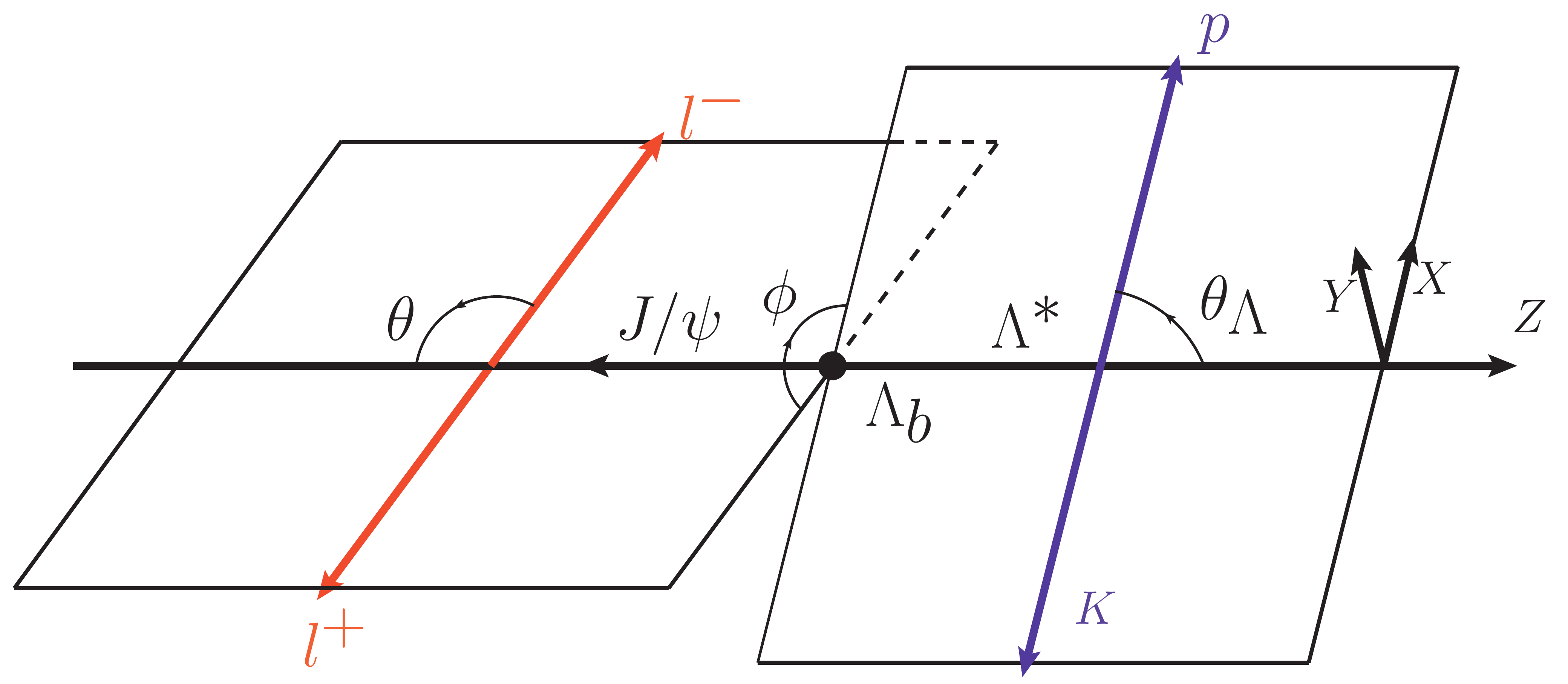} 
\caption{The kinematics for the  $\Lambda_b\to \Lambda^*_J(pK)J/\psi(\ell^+\ell^-)$ decay.  In the $\Lambda_b$ baryon  rest frame, the  $\Lambda^{*}_J$ moves along the $z$-axis. The $\theta(\theta_\Lambda)$ is defined as the angle between negative (positive) $z$-axis and the moving  direction of  $\ell^-$ ($p)$ in the $J/\psi$ ($\Lambda^{*}_J$) rest frame. The $\phi$ is the angle between  the $\Lambda^{*}_J$ and $J/\psi$ cascade decay planes.}
\label{fig:dynamic}
\end{figure}

The decay kinematics for  $\Lambda_b\to \Lambda^*_J(pK)J/\psi(\ell^+\ell^-)$ is shown in Fig.~\ref{fig:dynamic}.  In the $\Lambda_b$ baryon  rest frame, the  $\Lambda^{*}_J$ moves along the $z$-axis. The $\theta(\theta_\Lambda)$ is defined as the angle between negative (positive) $z$-axis and the moving  direction of  $\ell^-$ ($p)$ in the $J/\psi$ ($\Lambda^{*}_J$) rest frame. The $\phi$ is the angle between the $\Lambda^{*}_J$ and $J/\psi$ cascade decay planes.

Decay amplitude for the four-body decays can be divided into Lorentz-invariant hadronic part and leptonic matrix elements: 
\begin{eqnarray}
i\mathcal{M}(\Lambda_b\to \Lambda_J^*(pK) J/\psi(\ell^+\ell^-))
&=& \sum_{\Lambda^*_J}\sum_{s_{\Lambda_{J}^*} s_{J/\psi}}i \mathcal{M}(J/\psi\to\ell^+\ell^-)\frac{i}{q^2-m^2_{J/\psi}+im_{J/\psi}\Gamma_{J/\psi}} i\mathcal{M}(\Lambda_b\to\Lambda_J^*J/\psi) \nonumber\\
&&\times \frac{i}{p^2_{\Lambda_J^*}-m^2_{\Lambda^*_J}+im_{\Lambda^*_J}\Gamma_{\Lambda^*_J}} i \mathcal{M}(\Lambda^*_J\to pK),\label{amplitude}
\end{eqnarray}
with the $J/\psi$ momentum: $q^\mu= p_{\ell^+}^\mu+ p_{\ell^-}^\mu$, and the $\Lambda_{J}^*$ momentum: $p_{\Lambda_{J}^*}^\mu=p_p^\mu+p_K^\mu$. In the above expression, a resonance approximation has been adopted for the production of $pK^-$ and lepton pair.

Since the individual parts with a specific polarization are Lorentz invariant, they can be calculated in different reference frames. 
The  $\mathcal{M}(\Lambda_b\to \Lambda_J^* J/\psi)$ is induced by the $b\to s c\bar c$ transition whose effective Hamiltonian is: 
\begin{eqnarray}
\mathcal H_{eff}(b\to sc\bar c)=\frac{G_F}{\sqrt{2}}\bigg(V_{cb}V^*_{cs}(C_1O_1+C_2O_2)\bigg),
\end{eqnarray}
with
\begin{eqnarray}
O_1=[\bar c_\alpha\gamma^\mu(1-\gamma_5) b_\beta][\bar s_\beta\gamma_\mu(1-\gamma_5) c_\alpha],\quad O_2=[\bar c_\alpha\gamma^\mu(1-\gamma_5) b_\alpha][\bar s_\beta\gamma_\mu(1-\gamma_5) c_\beta].
\end{eqnarray} 
The $G_F$ and $V_{cb}, V_{cs}$ are Fermi coupling constant and   Cabibbo-Kobayashi-Maskawa matrix element, respectively. $O_i$ is the low-energy effective operator and $C_i$ is the corresponding Wilson coefficient obtained by integrating out high energy contributions. Applying the Fierz transformation and adopting the factorization ansatz, one can write the $\Lambda_b\to \Lambda^*_J J/\psi$ amplitude    as: 
\begin{eqnarray}
\mathcal{M}(\Lambda_b\to \Lambda_J^* J/\psi)&=&\frac{G_F}{\sqrt{2}}V_{cb}V^*_{cs} a_2f_{J/\psi}m_{J/\psi}\langle \Lambda_J^*|\bar s\gamma^\mu(1-\gamma_5)b|\Lambda_b\rangle \epsilon^{*}_\mu(s_{J/\psi}), \notag\\
a_2&=&C_1+C_2/N_c, ,\label{lambdabff}
\end{eqnarray} 
with $m_{J/\psi}=3.097$GeV, $m_{\Lambda_{b}}=5.619$GeV, $V_{cs}=0.975$, $V_{cb}=0.041$, $f_{J/\psi}=0.405$GeV. The $f_{J/\psi}$ and $N_{c}$ are  the decay constant of $J/\psi$ and the color number for quarks, respectively. The Wilson coefficients at $m_b$ scale are used as $C_{1}(m_b)=-0.248$ and $C_{2}(m_b)=1.107$~\cite{Buchalla:1995vs}.

The leptonic decay amplitude of $J/\psi$ can be calculated with  an effective Hamiltonian: 
\begin{eqnarray}
i\mathcal{M}(J/\psi\to \ell^+\ell^-)
&=&
\langle \ell^+(s_+)\ell^-(s_-)| -i gF^{\mu\nu} {F}^{\prime}_{\mu\nu}|J/\psi(s_{J/\psi})\rangle\nonumber
\\
&= & 2i eg \times\bar{u}(s_-)\gamma^\mu v(s_+)\epsilon_\mu(s_{J/\psi})\nonumber
\\
&= & 2i eg \times L^{s_{J/\psi}}_{s_-,s_+}(\theta, \phi), 
\end{eqnarray} 
where $s_\pm$ and $s_{J/\psi}$ are the helicity of the $\ell^\pm$ and $J/\psi$ respectively.
The $F_{\mu\nu}=\partial_\mu A_\nu-\partial_\nu A_\mu$ is the  electromagnetic field strength tensor and  ${F}^{\prime}_{\mu\nu}=\partial^{\mu}A_{J/\psi}^{\nu}-\partial^\nu A_{J/\psi}^{\mu}$ characterizes the  $J/\psi$. The explicit results for   $L^{s_{J/\psi}}_{s_-,s_+}(\theta, \phi)$ are given in the appendix. The coupling constant $g$ can be determined from the   $J/\psi$ leptonic decay width: 
\begin{eqnarray}
g^2=\frac{3\Gamma(J/\psi\to \ell^+\ell^-)m_{J/\psi}^2}{4\alpha_{em}(m_{J/\psi}^2+2m_\ell^2)\sqrt{m_{J/\psi}^2-4m_\ell^2}}.
\end{eqnarray}

The hadron decay $\Lambda_J^*\to pK^-$  is parametrized as  
\begin{eqnarray}
i\mathcal{M}(\Lambda^*_J\to pK)=\mathcal{A}_J\times (D^{J_{\Lambda_J^*}}_{s_{\Lambda_J^*},s_p}(\phi_\Lambda,\theta_{\Lambda}))^*,\;J=1520,1600,1800, \label{lambdatopk}
\end{eqnarray}
where $J_{\Lambda_J^*}$ is the total spin of the $\Lambda_J^*$,  and $s_{\Lambda_J^*}$ and $s_p$ are the helicities,  respectively.
The $D^{J_{\Lambda^*}}_{s_{\Lambda_J^*},s_p}(\phi_\Lambda,\theta_\Lambda)$ is Wigner function~\cite{Workman:2022}, whose explicit expression is also given in the appendix \ref{apA}. It should be noticed that the $\phi_\Lambda$ is the angle from the $\Lambda^*_J pK^-$ plane and the $x-z$ plane, and can be chosen as $0$ in the calculation. Eq.~\eqref{lambdatopk} applies to the distribution for any pertinent resonance, and in this analysis we consider the $\Lambda^*_{1520}$, $\Lambda^*_{1600}$ and $\Lambda^*_{1800}$. 
Using the two body decay process $\Lambda^*_J\to pK$, one can extract the coupling strength  $\mathcal{A}_J$ as 
\begin{eqnarray}
\mathcal{A}_J=\sqrt{\Gamma(\Lambda^*_J\to pK)16\pi m_{\Lambda_{J}^*}^2/|\vec p_p|},~~~J=1520\;,\nonumber \\
\mathcal{A}_J=\sqrt{\Gamma(\Lambda^*_J\to pK)8\pi m_{\Lambda_{J}^*}^2/|\vec p_p|},~~~J=1600,1800\;.
\end{eqnarray}

Then the decay amplitude of $\Lambda_b\to \Lambda^*_J(pK)J/\psi(\ell^+\ell^-)$ process is calculated  as:  
 \begin{eqnarray}
 i\mathcal{M}(\Lambda_b\to \Lambda^*_J(pK)J/\psi(\ell^+\ell^-)) &=& \frac{ge}{q^2-m^2_{J/\psi}+im_{J/\psi}\Gamma_{J/\psi}} L^{s_{J/\psi}}_{s_-,s_+}(\theta, \phi) \nonumber\\
 && \times  \frac{1}{M_{pK}^2-m^2_{\Lambda^*_J}+im_{\Lambda^*_J}\Gamma_{\Lambda^*_J}} \mathcal{A}_J  D^{*J_{\Lambda^*_J}}_{s_{\Lambda^*},s_p}(\phi_\Lambda=0,\theta_\Lambda)  \nonumber\\
 && \times i \mathcal{M}(\Lambda_b\to \Lambda^*_J J/\psi),\label{helicity}
 \end{eqnarray}  
 For the sake of simplicity, one can introduce the abbreviation  $\mathcal{A}^{s_{\Lambda_b}}_{s_p,s_{J/\psi}}(\theta_\Lambda)$ for the hadronic part: 
\begin{eqnarray} 
\mathcal{A}^{s_{\Lambda_b}}_{s_p,s_{J/\psi}}(\theta_\Lambda)&=& \sum_{J=\frac{1}{2},\frac{3}{2}}H^{ J_{\Lambda^*_J}}_{s_{\Lambda_b},s_{\Lambda^*_J}}\times (D^{J_{\Lambda^*_J}}_{s_{\Lambda^*_J},s_p}(0,\theta_\Lambda))^*,\notag\\
H^{\frac{3}{2}}_{s_{\Lambda_b},s_{\Lambda^*_J}}&=&L_{\Lambda^*_{1520}}(M_{pK}^2,m_{\Lambda_J^*}) i\mathcal{M}(\Lambda_b\to \Lambda^*_{1520} J/\psi),\notag\\
H^{\frac{1}{2}}_{s_{\Lambda_b},s_{\Lambda^*_J}}&=& L_{\Lambda^*_{1600}}(M_{pK}^2,m_{\Lambda^*_J}) i\mathcal{M}(\Lambda_b\to \Lambda^*_{1600} J/\psi)\notag\\
&& + L_{\Lambda^*_{1800}}(M_{pK}^2,m_{\Lambda^*_J})i \mathcal{M}(\Lambda_b\to \Lambda^*_{1800} J/\psi), \nonumber\\\label{amp1}
L_{\Lambda^*_J}(M_{pK}^2,m_{\Lambda^*_J})&=&\mathcal{A}_J\frac{1}{M_{pK}^2-m^2_{\Lambda^*_J}+im_{\Lambda^*_J}\Gamma_{\Lambda^*_J}}.
\end{eqnarray}

The differential decay width is formulated as
\begin{eqnarray}
d\Gamma=d\Pi_4\times\frac{(2\pi)^4}{2m_{\Lambda_b}}|\mathcal M(\Lambda_b\to\Lambda_J^*(pK)J/\psi(\ell^+\ell^-))|^2,
\end{eqnarray} 
where the phase space is used as
\begin{align}
d\Pi_4(p_{\ell^+},p_{\ell^-},p_{p},p_{K})=&(2\pi)^3(2\pi)^3dq^2dM_{pK}^2\times d\Pi_2(p_{\ell^+},p_{\ell^-})\times d\Pi_2(p_{p},p_{K})\times d\Pi_2(p_{J/\psi},p_{\Lambda_J^*})\nonumber \\
=&\frac{\sqrt{\lambda(m_{\Lambda_{b}},m_{\Lambda_J^*},q^2)}|\vec{p}_{p}|\sqrt{q^2-4m_\ell^2}}{(2\pi)^{10} \times 128\sqrt{q^2} m_{\Lambda_{b}}^2\sqrt{M_{pK}^2}}d\cos\theta d\cos\theta_{\Lambda} d\phi dM_{pK}^2 dq^2,
\end{align} 
with $|\vec p_{p}|=\sqrt{\lambda(m_{\Lambda^{*}_J},m_{K},m_{p})}/(2m_{\Lambda^{*}_J})$, $\lambda(m_{\Lambda^{*}_J},m_{K},m_{p})=((m_{\Lambda^{*}_J}+m_{K})^{2}-m_{p}^{2})((m_{\Lambda^{*}_J}-m_{K})^{2}-m_{p}^{2})$ and $M_{pK}^2=p_{\Lambda^*_J}^2$.

\section{ angular distribution of $\Lambda_b\to\Lambda^*_J(pK)J/\psi(\ell^+\ell^-)$}

Combining all the  elements, one   obtains the differential decay width for the four body decay process $\Lambda_b\to\Lambda_{J}^*(pK)J/\psi(\ell^+\ell^-))$ as 
\begin{eqnarray}
\frac{d\Gamma(\Lambda_b\to\Lambda_{J}^*(pK)J/\psi(\ell^+\ell^-))}{d\cos\theta d\cos\theta_\Lambda d\phi dq^2 dM_{pK}^2}&=&\frac{\sqrt{\lambda(m_{\Lambda_{b}},m_{\Lambda^*_J},q^2)}|\vec{p}_{p}|}{8192\pi^6 m_{\Lambda_{b}}^3m_{\Lambda^*_J}}\notag\\
&&\times\frac{1}{2}\sum_{s_{\Lambda_b},s_p,s_+,s_-}\frac{3\pi \Gamma(J/\psi\to \ell^+\ell^-)m_{J/\psi}|L^{s_{J/\psi}}_{s_-,s_+}(\phi,\theta)\mathcal{A}^{s_{\Lambda_b}}_{s_p,s_{J/\psi}}(\theta_\Lambda)|^2}{(m_{J/\psi}^2+2m_\ell^2)|q^2-m_{J/\psi}^2+ i m_{J/\psi}\Gamma_{J/\psi}|^2}.
\end{eqnarray} 
Using the narrow-width limit for the $J/\psi$
\begin{eqnarray}
 \frac{\Gamma_{J/\psi} m_{J/\psi} }{|(q^2-m_{J/\psi}^2) + i m_{J/\psi}\Gamma_{J/\psi}|^2} = \pi  \delta(q^2-m_{J/\psi}^2),
\end{eqnarray}
one can  arrive at the differential decay width as 
\begin{eqnarray}
\frac{d\Gamma(\Lambda_b\to\Lambda_{J}^*(pK)J/\psi(\ell^+\ell^-))}{d\cos\theta d\cos\theta_\Lambda d\phi dM_{pK}^2}&=&\frac{3\sqrt{\lambda(m_{\Lambda_{b}},m_{\Lambda^*_J},m_{J/\psi})}|\vec{p}_{p}|}{16384\pi^4 m_{\Lambda_{b}}^3m_{\Lambda^*_J}(m_{J/\psi}^2+2m_\ell^2)}\notag\\
&&\times\frac{1}{2}\sum_{s_{\Lambda_b},s_p,s_+,s_-}\mathcal{B}(J/\psi\to \ell^+\ell^-) |L^{s_{J/\psi}}_{s_-,s_+}(\phi,\theta)\mathcal{A}^{s_{\Lambda_b}}_{s_p,s_{J/\psi}}(\theta_\Lambda)|^2.
\end{eqnarray} 
With the explicit expressions for $|L^{s_{J/\psi}}_{s_-,s_+}(\phi,\theta)\mathcal{A}^{s_{\Lambda_b}}_{s_p,s_{J/\psi}}(\theta_\Lambda)|^2$ given in the appendix, the angular distribution is derived as 
\begin{eqnarray}
\frac{d\Gamma(\Lambda_b\to\Lambda_{J}^*(pK)J/\psi(\ell^+\ell^-))}{d\cos\theta d\cos\theta_\Lambda d\phi dM_{pK}^2}&=&\mathcal{P} \bigg(L_1+L_{2}\cos2\phi+L_3\cos2\theta+ L_{4}\sin2\theta\cos\phi+L_5\cos2\phi\cos2\theta+\notag\\
&&L_6\sin2\theta\sin\phi+L_7\sin2\phi+L_8\cos2\theta\sin2\phi\bigg),\notag\\
\mathcal{P}&=&\frac{3\sqrt{\lambda(m_{\Lambda_{b}},m_{\Lambda^*_J},m_{J/\psi})}|\vec{p}_{p}|}{ 8192\pi^4 m_{\Lambda_{b}}^3m_{\Lambda^*_J}(1+2\hat{m}_\ell^2)}\mathcal{B}(J/\psi\to \ell^+\ell^-).\label{dw}
\end{eqnarray} 
The angular coefficients $L_i$($i=1-8$) are given as  
\begin{align}
L_1=&\sum_{s_{\Lambda_b},s_p}\bigg(2\hat{m}_\ell^2(|\mathcal{A}^{s_{\Lambda_b}}_{s_p,-1}(\theta_\Lambda)|^2+2|\mathcal{A}^{s_{\Lambda_b}}_{s_p,0}(\theta_\Lambda)|^2+|\mathcal{A}^{s_{\Lambda_b}}_{s_p,1}(\theta_\Lambda)|^2)+\frac{1}{2}(3|\mathcal{A}^{s_{\Lambda_b}}_{s_p,-1}(\theta_\Lambda)|^2+2|\mathcal{A}^{s_{\Lambda_b}}_{s_p,0}(\theta_\Lambda)|^2+3|\mathcal{A}^{s_{\Lambda_b}}_{s_p,1}(\theta_\Lambda)|^2)\bigg),\notag\\
L_2=&-(4\hat{m}_\ell^2-1)\sum_{s_{\Lambda_b},s_p}\mathcal{R}_e(\mathcal{A}^{s_{\Lambda_b}}_{s_p,-1}(\theta_\Lambda)\mathcal{A}^{s_{\Lambda_b}*}_{s_p,1}(\theta_\Lambda)),\notag\\
L_3=&\frac{-1}{2}(4\hat{m}_\ell^2-1)\sum_{s_{\Lambda_b},s_p}(|\mathcal{A}^{s_{\Lambda_b}}_{s_p,-1}(\theta_\Lambda)|^2-2|\mathcal{A}^{s_{\Lambda_b}}_{s_p,0}(\theta_\Lambda)|^2+|\mathcal{A}^{s_{\Lambda_b}}_{s_p,1}(\theta_\Lambda)|^2),\notag\\
L_4=&-\sqrt{2}(4\hat{m}_\ell^2-1)\sum_{s_{\Lambda_b},s_p}\mathcal{R}_e(\mathcal{A}^{s_{\Lambda_b}}_{s_p,0}(\theta_\Lambda)(\mathcal{A}^{s_{\Lambda_b}*}_{s_p,-1}(\theta_\Lambda)-\mathcal{A}^{s_{\Lambda_b}*}_{s_p,1}(\theta_\Lambda))),\notag\\
L_5=&(4\hat{m}_\ell^2-1)\sum_{s_{\Lambda_b},s_p}\mathcal{R}_e(\mathcal{A}^{s_{\Lambda_b}}_{s_p,-1}(\theta_\Lambda)\mathcal{A}^{s_{\Lambda_b}*}_{s_p,1}(\theta_\Lambda)),\notag\\
L_6=&-\sqrt{2}(4\hat{m}_\ell^2-1)\sum_{s_{\Lambda_b},s_p}\mathcal{I}_m(\mathcal{A}^{s_{\Lambda_b}}_{s_p,0}(\theta_\Lambda)(\mathcal{A}^{s_{\Lambda_b}*}_{s_p,-1}(\theta_\Lambda)+\mathcal{A}^{s_{\Lambda_b}*}_{s_p,1}(\theta_\Lambda))),\notag\\
L_7=&(4\hat{m}_\ell^2-1)\sum_{s_{\Lambda_b},s_p}\mathcal{I}_m(\mathcal{A}^{s_{\Lambda_b}}_{s_p,-1}(\theta_\Lambda)\mathcal{A}^{s_{\Lambda_b}*}_{s_p,1}(\theta_\Lambda))=-L_8.
\end{align}
Then one can explore the $L_{i}(i=1-8)$ by expanding $\mathcal{A}^{s_{\Lambda_b}}_{s_p,s_{J/\psi}}$ which contain  the resonance of $\Lambda_{1520,1600,1800}^*$. The specific expression including $\theta_\Lambda$ can be displayed in appendix~\ref{coef}.  
Thus the differential decay width for $\Lambda_{b} \to \Lambda_{J}^{*}( p K^{-}) J/\psi( \ell^{+}\ell^{-})$ as a function of $\theta_{\Lambda}$, $\theta$, $\phi$ and $M_{pK}^{2}$ is given as
\begin{eqnarray}
\frac{d\Gamma(\Lambda_b\to\Lambda_{J}^*(pK)J/\psi(\ell^+\ell^-))}{d\cos\theta d\cos\theta_\Lambda d\phi dM_{pK}^2}&=&\mathcal{P}\bigg(L_{11}+\cos\theta_\Lambda L_{12}+\cos2\theta_\Lambda L_{13}+\cos2\phi(L_{21}+\cos2\theta_\Lambda L_{22})\notag\\
&&+ \cos2\theta(L_{31}+\cos\theta_\Lambda L_{32}+\cos2\theta_\Lambda L_{33})+ \sin2\theta\cos\phi(\sin\theta_\Lambda L_{41}+\sin2\theta_\Lambda L_{42})\notag\\
&&+\cos2\phi\cos2\theta(L_{51}+\cos2\theta_\Lambda L_{52})+\sin2\theta\sin\phi(\sin\theta_\Lambda L_{61}+\sin2\theta_{\Lambda} L_{62})\notag\\
&&+\sin2\phi(L_{71}+\cos2\theta_\Lambda L_{72})+\cos2\theta\sin2\phi(L_{81}+\cos2\theta_\Lambda L_{82})\bigg).\label{dwa}
\end{eqnarray}
Here the formulas of $L_{ij}(i=1-8,j=1-3)$ are also given in appendix~\ref{coef}.


\section{Phenomenological applications}
\subsection{Transition Form Factors}

The hadron matrix element $\langle \Lambda^*_J|\bar s\gamma^\mu(1-\gamma_5)b|\Lambda_b\rangle$ in Eq.~\eqref{lambdabff} can be parameterized by form factors. For the  $\Lambda_b\to\Lambda_{1520}^*$ transition, one can define the  helicity-based form factors as~\cite{Meinel:2021mdj}:
\begin{eqnarray}
\langle \Lambda^*_{1520}(p^\prime, s^\prime)|\bar s\gamma^\mu b|\Lambda_b(p, s)\rangle&=&\bar u_\lambda(p^\prime, s^\prime)\bigg(f^{3/2}_0\frac{m_{\Lambda_{1520}^*}}{s_{p+}}\frac{(m_{\Lambda_b}-m_{\Lambda_{1520}^*})p^\lambda q^\mu}{q^2}\notag\\
&&+f^{3/2}_+\frac{m_{\Lambda_{1520}^*}}{s_{p-}}\frac{(m_{\Lambda_b}+m_{\Lambda_{1520}^*})p^\lambda (q^2(p^\mu+p^{\prime\mu})-q^\mu(m^2_{\Lambda_b}-m^2_{\Lambda_{1520}^*}))}{q^2 s_{p+}}\notag\\
&&+f^{3/2}_\perp\frac{m_{\Lambda_{1520}^*}}{s_{p-}}(p^\lambda\gamma^\mu-\frac{2p^\lambda(m_{\Lambda_b}p^{\prime\mu}+m_{\Lambda_{1520}^*}p^\mu)}{s_{p+}})\notag\\
&&+f^{3/2}_{\perp\prime}\frac{m_{\Lambda_{1520}^*}}{s_{p-}}(p^\lambda\gamma^\mu-\frac{2p^\lambda p^{\prime\mu}}{m_{\Lambda_{1520}^*}}+\frac{2p^\lambda(m_{\Lambda_b}p^{\prime\mu}+m_{\Lambda_{1520}^*}p^\mu)}{s_{p+}}+\frac{s_{p-}g^{\lambda\mu}}{m_{\Lambda_{1520}^*}})\bigg)u(p, s),\notag
\end{eqnarray}
\begin{eqnarray}
\langle \Lambda^*_{1520}(p^\prime, s^\prime)|\bar s\gamma^\mu \gamma_5 b|\Lambda_b(p, s)\rangle&=&\bar u_\lambda(p^\prime, s^\prime)\bigg(-g^{3/2}_0\gamma_5\frac{m_{\Lambda_{1520}^*}}{s_{p-}}\frac{(m_{\Lambda_b}+m_{\Lambda_{1520}^*})p^\lambda q^\mu}{q^2}\notag\\
&&-g^{3/2}_+\gamma_5\frac{m_{\Lambda_{1520}^*}}{s_{p+}}\frac{(m_{\Lambda_b}-m_{\Lambda_{1520}^*})p^\lambda (q^2(p^\mu+p^{\prime\mu})-q^\mu(m^2_{\Lambda_b}-m^2_{\Lambda_{1520}^*}))}{q^2 s_{p-}}\notag\\
&&-g^{3/2}_\perp\gamma_5\frac{m_{\Lambda_{1520}^*}}{s_{p+}}(p^\lambda\gamma^\mu-\frac{2p^\lambda(m_{\Lambda_b}p^{\prime\mu}-m_{\Lambda_{1520}^*}p^\mu)}{s_{p-}})\notag\\
&&-g^{3/2}_{\perp\prime}\gamma_5\frac{m_{\Lambda_{1520}^*}}{s_{p+}}(p^\lambda\gamma^\mu+\frac{2p^\lambda p^{\prime\mu}}{m_{\Lambda_{1520}^*}}+\frac{2p^\lambda(m_{\Lambda_b}p^{\prime\mu}+m_{\Lambda_{1520}^*}p^\mu)}{s_{p-}}-\frac{s_{p+}g^{\lambda\mu}}{m_{\Lambda_{1520}^*}})\bigg)u(p, s),\label{1520}
\end{eqnarray}
with $q^\mu=p^\mu-p^{\prime\mu}$ being  the transfered   momentum   and $s_{p\pm}=(m_{\Lambda_b}\pm m_{\Lambda_J^*})^2-q^2$, $q^2=m_{J/\psi}^2$.

These form factors have been calculated from Lattice QCD (LQCD)~\cite{Meinel:2021mdj}, where multi sets of lattice ensembles are used. To access the $M_{pK}^2$ distributions, the form factors are parametrized as~\cite{Meinel:2021mdj}
\begin{align}
f(M_{pK}^2)=F\left[1+C\frac{m_{\pi}^{2}-m_{\pi,phys}^{2}}{(4\pi f_{\pi})^2}+D a^{2} \Lambda^{2}\right]+A\left[1+C^{\prime}\frac{m_{\pi}^{2}-m_{\pi,phys}^{2}}{(4\pi f_{\pi})^2}+D^{\prime} a^{2} \Lambda^{2}\right](\omega-1),
\end{align} 
where the parameters F, A, C, D,$C^{\prime}$,$D^{\prime}$ are fitted from the lattice data and $\omega=(m_{\Lambda_b}^2+M_{pK}^2-m_{J/\psi}^2)/2m_{\Lambda_b}m_{\Lambda^*_J}$.  In the LQCD calculation, the finite lattice spacing and pion mass effects are also considered.  In the physical pion limit, $m_{\pi}=135$~MeV, and the continuum limit $a=0$, and using the $f_{\pi}=132$ MeV, $\Lambda=300$ MeV, one can  simplify the above parametrization as
\begin{eqnarray}
f(M_{pK}^2)=F+A(\omega-1).\label{qmm}
\end{eqnarray} 
 For the $\Lambda_{b}\to\Lambda^{*}_{1520}$ transition, results for the inputs  F and A are shown in Table~\ref{table1}, and in the following we will use these results as default.

If the final baryon is a spin-$\frac{1}{2}$ hadron, the weak transition form factor is parametrized as~\cite{Mott:2011cx}:
\begin{eqnarray}
\langle \Lambda^*_J(p^\prime, s^\prime)|\bar{s}\gamma^\mu b|\Lambda_b(p, s)\rangle&=&\bar{u}(p^\prime, s^\prime)\big(\gamma_\mu f^{p}_1+\frac{p_{\Lambda_b}^\mu}{m_{\Lambda_b}}f^p_2+\frac{p_{\Lambda_J^*}^\mu}{m_{\Lambda^*_J}}f^p_3\big)u(p, s),\notag\\
\langle \Lambda^*_J(p^\prime, s^\prime)|\bar{s}\gamma^\mu\gamma_5 b|\Lambda_b(p, s)\rangle&=&\bar{u}(p^\prime, s^\prime)\big(\gamma_\mu g^{p}_1+\frac{p_{\Lambda_b}^\mu}{m_{\Lambda_b}}g^p_2+\frac{p_{\Lambda^*_J}^\mu}{m_{\Lambda^*_J}}g^p_3\big)\gamma_5u(p, s).\label{12}
\end{eqnarray}
In Ref.~\cite{Mott:2011cx},  a model with a full quark model wave function and the full relativistic form of the quark is adopted to investigate the form factors, and  these form factors are studied in multi-component numerical (MCN) model. The $M^2_{pK}$-dependence  is parameterized as
\begin{eqnarray}
f(M_{pK}^2)=(a_0+a_2 p_{\Lambda}^2+a_4 p_{\Lambda}^4)\exp\bigg(-\frac{6m_q^2 p_\Lambda^2}{2\tilde{m_\Lambda}^2(\alpha_{\Lambda_b}^2+\alpha_{\Lambda^*}^2)}\bigg). \label{lam}
\end{eqnarray}
Here $p_{\Lambda}$ represents one of the daughter baryon momentum in the $\Lambda_{b}$ rest frame.   The MCN model parameters $a_{0}$, $a_{2}$, and $a_{4}$ are given in Table.~\ref{table1} and Table~\ref{table2} respectively. Due to the lack of results for the $\Lambda_{b}\to\Lambda^{*}_{1800}$ transition, we use  the results for the $\Lambda_{b}\to\Lambda^{*}_{1405}$.  This may induce sizable uncertainties, and future detailed analysis can resolve this approximation.

\begin{table}[htbp!]
\caption{Input parameters in Eq.\eqref{qmm} and Eq.\eqref{lam} for $\Lambda^*_{1520}$.}\label{table1}\begin{tabular}{|c|c|c|c|c|c|c|c}\hline\hline
\multicolumn{3}{|c|}{ Lattic QCD}&\multicolumn{4}{|c|}{ MCN quark model}\\\hline \hline
form factor &F&  A&form factor&$a_0$&$a_2$&$a_4$\\\hline 
$f^{3/2}_0$&$3.54(29) $&$-14.7(3.3)$&$f_1$&-1.66&-0.295&0.00924\\\hline
$f^{3/2}_+$&$0.0432(64)$&$1.63(19)$&$f_2$&0.544&0.194&-0.00420\\\hline
$f^{3/2}_\perp$&$-0.068(18)$&$2.49(35)$&$f_3$&0.126&0.00799&-0.000365\\\hline
$f^{3/2}_{\perp\prime}$&$0.0461(18)$&$-0.161(27)$&$f_4$&-0.0330&-0.00977&0.00211\\\hline
$g^{3/2}_0$&$0.0024(38)$&$1.58(17)$&$g_1$&-0.964&-0.100&0.00264\\\hline
$g^{3/2}_+$&$2.95(25)$&$-12.2(2.9)$&$g_2$&0.625&0.219&-0.00508\\\hline
$g^{3/2}_\perp$&$2.92(24)$&$-11.8(2.8)$&$g_3$&-0.183&-0.0380&0.00351\\\hline
$g^{3/2}_{\perp\prime}$&$-0.037(14)$&$0.09(25)$&$g_4$&0.0530&0.0161&-0.00221\\\hline
 &&& $\alpha_{\Lambda_b}=0.443$&$\alpha_{\Lambda*(1520)}=0.333$&$\tilde{m_\Lambda}=1.1249$&$m_q=0.2848$\\\hline
\end{tabular}
\end{table}
\begin{table}[htbp!]
\caption{Input parameters in Eq.\eqref{qmm} and Eq.\eqref{lam} for spin-$1/2$ resonance $\Lambda^*_{1600,1800}$ in MCN quark model. }\label{table2}\begin{tabular}{|c|c|c|c|c|c|c|c|}\hline\hline
\multicolumn{4}{|c|}{ $\Lambda^*_{1600}$}&\multicolumn{4}{|c|}{ $\Lambda^*_{1800}$}\\\hline \hline
form factor&$a_0$&$a_2$&$a_4$&form factor&$a_0$&$a_2$&$a_4$\\\hline
$f^{+}_1$&0.467&0.615&0.0568&$f^{-}_1$&0.246&0.238&0.00976\\\hline
$f^{+}_2$&-0.381&-0.2815&-0.0399&$f^{-}_2$&-0.984&-0.0257&0.0173\\\hline
$f^{+}_3$&0.0501&-0.0295&-0.00163&$f^{-}_3$&0.118&0.0237&-0.000692\\\hline
$g^{+}_1$&0.114&0.300&0.0206&$g^{-}_1$&1.15&0.260&-0.00303\\\hline
$g^{+}_2$&-0.394&-0.307&-0.0445&$g^{-}_2$&-0.874&-0.0264&0.0159\\\hline
$g^{+}_3$&-0.0433&0.0478&0.00566&$g^{-}_3$&0.00871&-0.0196&-0.000997\\\hline
\multicolumn{4}{|c|}{$\alpha_{\Lambda*(1600)}=0.387$}&\multicolumn{4}{|c|}{$\alpha_{\Lambda*(1800)}=0.333$}\\\hline
\end{tabular}
\end{table}

\subsection{Numerical Results}

Two-body decays $\Lambda_b\to \Lambda_J^* J/\psi$ can provide a calibration for the four-body decay process, and the decay widths for $\Lambda_b\to \Lambda_J^* J/\psi$ are given  as 
\begin{eqnarray}
\Gamma(\Lambda_b\to \Lambda_{J}^* J/\psi)=\sum_{s_{\Lambda_b},s_{\Lambda_J^*},s_{J/\psi}}\frac{ |\vec{p_{\Lambda_J^*}}|}{8\pi m^2_{\Lambda_b}}\frac{1}{2}|\mathcal{M}(\Lambda_b\to \Lambda^*_J J/\psi)|^2.
\end{eqnarray} 
 With the form factors  from Ref.~\cite{Mott:2011cx},  one can calculate branching fractions for the process involving    different resonances $\Lambda_{1520,1600,1800}^{*}$: 
\begin{eqnarray}
\mathcal{B}(\Lambda_b\to \Lambda_{1520}^* J/\psi)&=&5.78\times 10^{-4},\notag\\
\mathcal{B}(\Lambda_b\to \Lambda_{1600}^* J/\psi)&=&2.44\times 10^{-4},\notag\\
\mathcal{B}(\Lambda_b\to \Lambda_{1800}^* J/\psi)&=&4.48\times 10^{-4}.  \label{eq:Lambdab_Lambda_Jpsi}
\end{eqnarray} 
There is no experimental measurement of the above three processes. However the available data indicates $\mathcal{B}(\Lambda_b\to J/\psi \Lambda)\times\mathcal{B}(b\to\Lambda_b)=(5.8\pm0.8)\times10^{-5}$\cite{D0:2011pqa,CDF:1996rvy}, where $\Lambda$ is the ground state. Using the estimate of the fragmentation fraction $\mathcal{B}(b\to\Lambda_b)= 0.175\pm0.106$~\cite{Hsiao:2015txa}, one can obtain: $\mathcal{B}(\Lambda_b\to J/\psi \Lambda)=(3.3\pm0.5\pm2.0)\times10^{-4}$,  which is at the same order   with the results in Eq.~\eqref{eq:Lambdab_Lambda_Jpsi}.

Based on the differential decay width  in Eq.~\eqref{dw}, one can obtain the differential  decay width: 
\begin{eqnarray} 
d\Gamma(\Lambda_b\to\Lambda_{J}^*(pK)J/\psi(\ell^+\ell^-))/dM_{pK}^2&=&\mathcal{P}
\frac{8\pi}{9}\bigg(9L_{11}-3L_{13}-3L_{31}+L_{33}\bigg).\label{dww}
\end{eqnarray} 
Using the inputs  from PDG~\cite{Workman:2022}, 
\begin{eqnarray}
&&\mathcal{B}(J/\psi\to e^+e^-)=(5.971\pm0.032)\%,\quad \mathcal{B}(J/\psi\to \mu^+\mu^-)=(5.961\pm0.033)\%,\notag\\&&\mathcal{B}(\Lambda_{1520}^*\to p K^-)=(22.5\pm0.5)\%,\quad\mathcal{B}(\Lambda_{1600}^*\to p K^-)\sim(10.5)\%,\quad
\mathcal{B}(\Lambda_{1800}^*\to p K^-)\sim(16)\%,\notag\\
&&\Gamma_{1520}=(0.01573\pm0.00026)\rm{GeV},\Gamma_{1600}\sim\Gamma_{1800}\sim 0.2\rm{GeV}, m_{p}=0.938{GeV}, m_{K}=0.494{GeV},
\end{eqnarray} 
one can obtain the $\Lambda_{b}$ four-body decay widths with final state $p K$ produced by a determined resonance $\Lambda^*_{1520,1600,1800}$ as
\begin{eqnarray} 
&&\mathcal{B}(\Lambda_b\to\Lambda^*(pK)J/\psi(\mu^+\mu^-))=(1.35\pm0.28)\times 10^{-5},\quad\;\;\mathcal{B}(\Lambda_b\to\Lambda^*(pK)J/\psi(e^+e^-))=(1.35\pm0.28)\times 10^{-5},\notag\\
&&\mathcal{B}(\Lambda_b\to\Lambda_{1520}^*(pK)J/\psi(\mu^+\mu^-))=(7.22\pm2.53)\times 10^{-6},\;\mathcal{B}(\Lambda_b\to\Lambda_{1520}^*(pK)J/\psi(e^+e^-))=(7.22\pm2.54)\times 10^{-6},\notag\\
&&\mathcal{B}(\Lambda_b\to\Lambda_{1600}^*(pK)J/\psi(\mu^+\mu^-))=1.11\times 10^{-6},\;\quad\quad\quad\mathcal{B}(\Lambda_b\to\Lambda_{1600}^*(pK)J/\psi(e^+e^-))=1.11\times 10^{-6},\notag\\
&&\mathcal{B}(\Lambda_b\to\Lambda_{1800}^*(pK)J/\psi(\mu^+\mu^-))=3.87\times 10^{-6},\;\quad\quad\quad\mathcal{B}(\Lambda_b\to\Lambda_{1800}^*(pK)J/\psi(e^+e^-))=3.88\times 10^{-6}.\label{Br}
\end{eqnarray}
The $\Lambda_b\to \Lambda^*_{1600,1800}$ form factors  are used from the MCN model~\cite{Mott:2011cx},  and  no uncertainties are given.  It is interesting to notice that such results are also in agreement with the results for two-body decays in the narrow width approximation.

\begin{figure}[htbp!]
  \begin{minipage}[t]{0.4\linewidth}
  \centering
  \includegraphics[width=1.0\columnwidth]{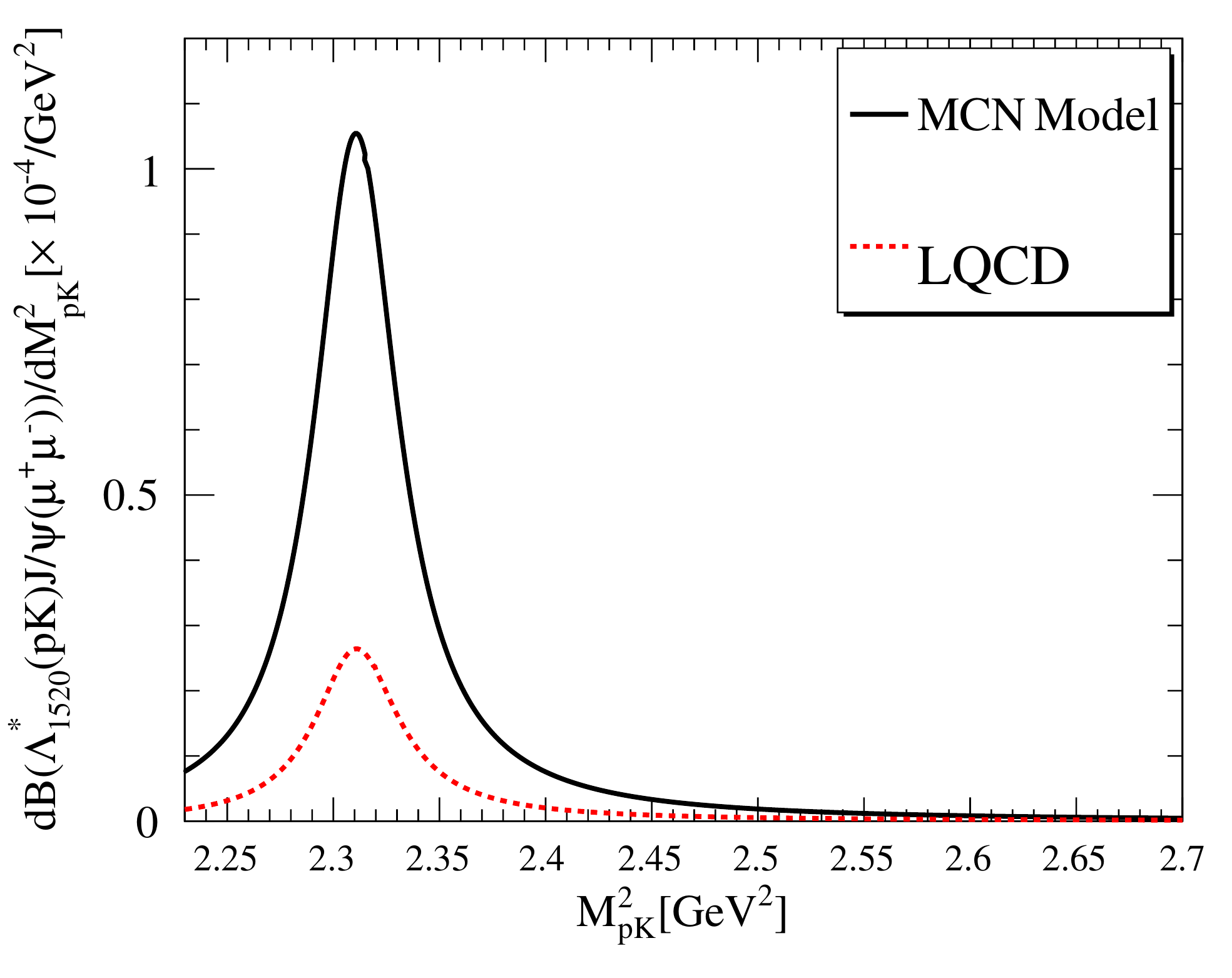}
    \end{minipage}
 \caption{The differential branching fraction $d\mathcal{B}/dM_{pK}^2$ for the process $\Lambda_b\to\Lambda_{1520}^*(pK)J/\psi(\ell^+\ell^-),\ell=\mu$ (in units of $10^{-4}/{\rm GeV}^2$)  with Lattice QCD~\cite{Meinel:2021mdj} and the MCN quark model~\cite{Mott:2011cx} form factors.}  \label{LAQM}
\end{figure}

If the MCN model results for the $\Lambda_b\to\Lambda_{1520}^*$ form factors  are used, we can find $\mathcal{B}(\Lambda_b\to\Lambda_{1520}^*(pK)J/\psi(\mu^+\mu^-))=1.904\times10^{-6}$, which is reduced by a factor of ~3. In  Fig.~\ref{LAQM}, we show the differential decay branching fraction $d\mathcal{B}/dq^2(\Lambda_b\to\Lambda_{1520}^*(pK)J/\psi(\ell^+\ell^-))$ with the two sets of form factors. It can be seen that a significant discrepancy appears at the low-$M_{pK}^2$ region for different forms of parameterized form factors.

\begin{figure}[htbp!]
  \begin{minipage}[t]{0.4\linewidth}
  \centering
  \includegraphics[width=1.0\columnwidth]{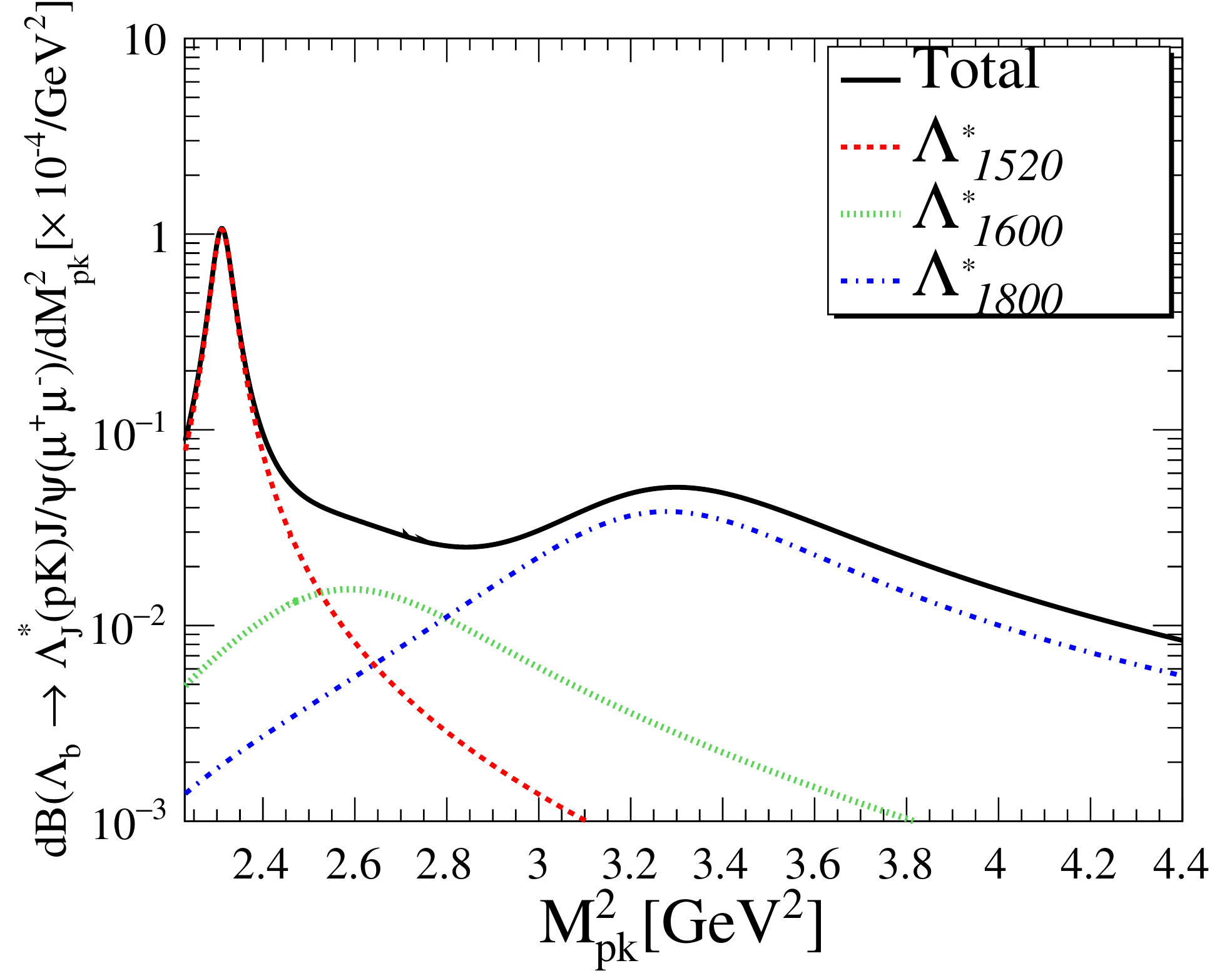}
    \end{minipage}
    \begin{minipage}[t]{0.4\linewidth}
  \centering
  \includegraphics[width=1.0\columnwidth]{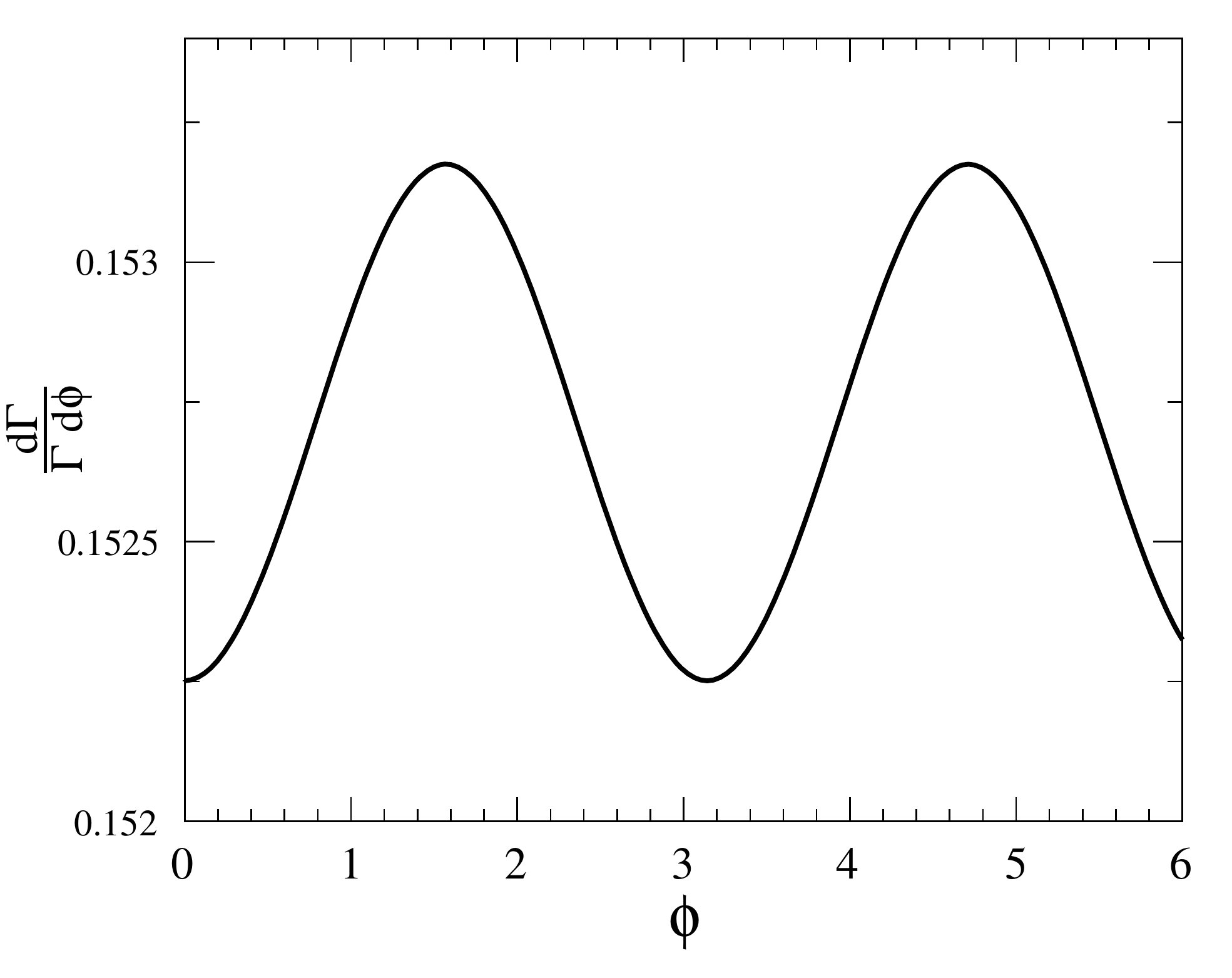}
    \end{minipage}
    \begin{minipage}[t]{0.4\linewidth}
  \centering
  \includegraphics[width=1.0\columnwidth]{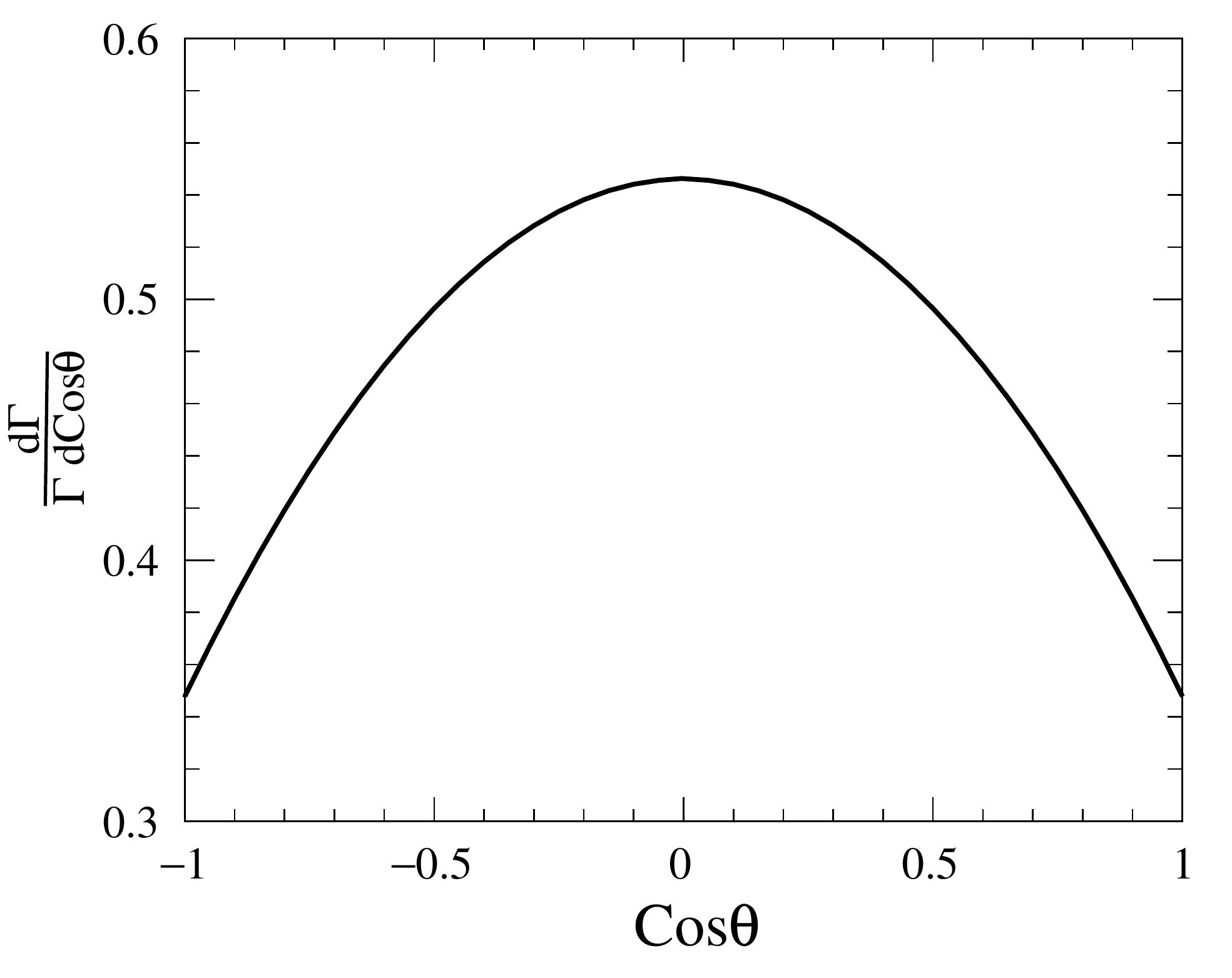}
    \end{minipage}
    \begin{minipage}[t]{0.4\linewidth}
  \centering
  \includegraphics[width=1.0\columnwidth]{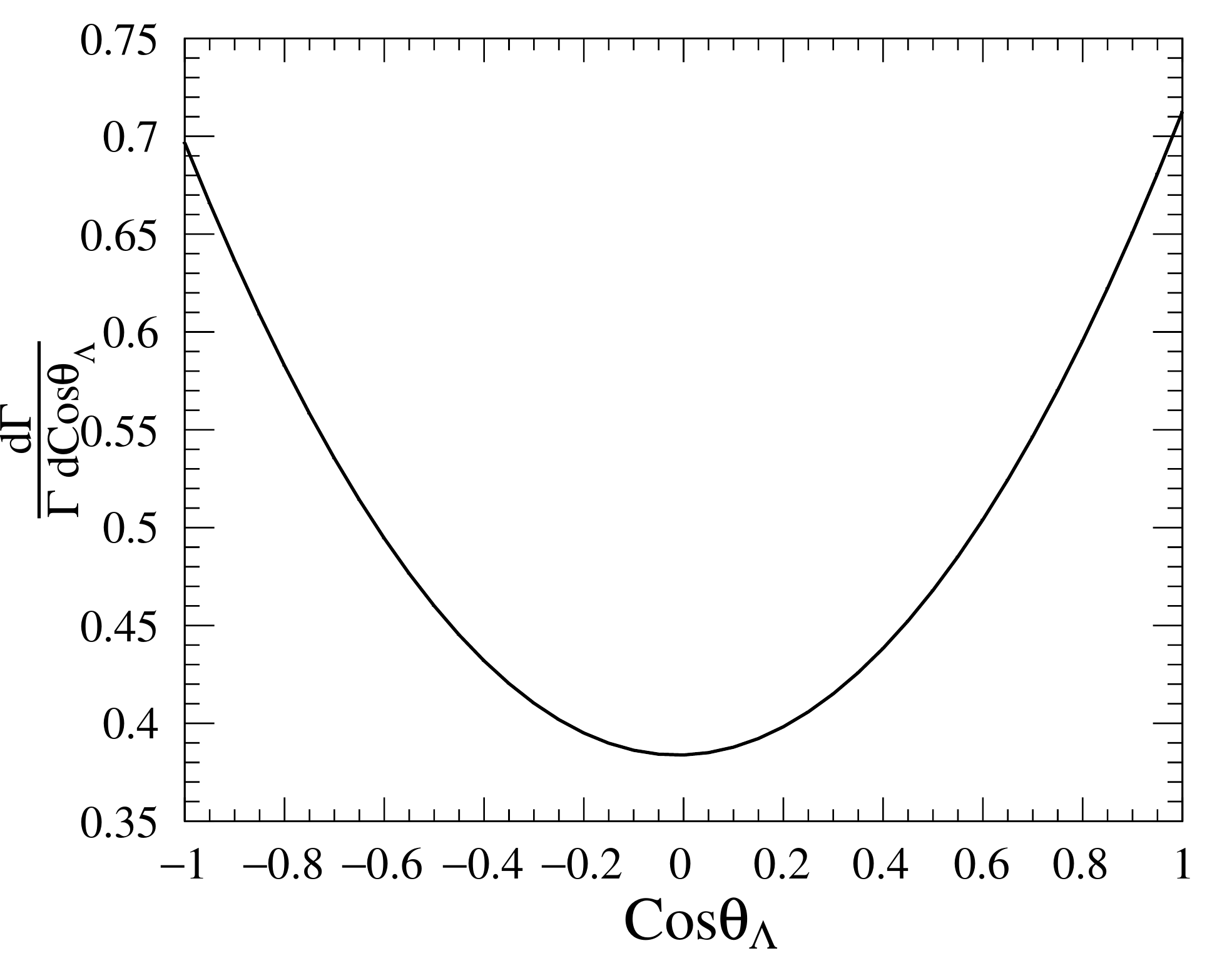}
    \end{minipage}
 \caption{The ($d\mathcal{B}/dM_{pK}^2$, $\frac{d\Gamma}{\Gamma d\phi}$, $\frac{d\Gamma}{\Gamma d\cos\theta}$, $\frac{d\Gamma}{\Gamma d\cos\theta_\Lambda}$) of process $\Lambda_b\to\Lambda^*(pK)J/\psi(\mu^+\mu^-)$.}  \label{decay width}
\end{figure}

The differential decay widths for the processes $\Lambda_{b}\to\Lambda_{J}^{*}(pK^{-})J/\psi(\ell^{+}\ell^{-})$ as a function of $M_{pK}^{2}$ are given in Fig.~\ref{decay width}.  We also show the normalized $\phi$ angular distribution for the $\Lambda_{b}$ decay in Fig.~\ref{decay width}.  
Since the lepton pair arises from the decay of $J/\psi$ induced by vector current,  angular distributions   for the lepton are proportional to $\cos2\theta$.

\subsubsection{Distribution of $\theta_\Lambda$ }

One can integrate the angle $\theta,\phi$  and  explore the normalized distribution of $\theta_{\Lambda}$,
\begin{eqnarray} 
\frac{1}{\Gamma}d\Gamma(\Lambda_b\to\Lambda_J^*(pK)J/\psi(\ell^+\ell^-))/dM_{pK}^2d\cos\theta_\Lambda&=&\bigg(L_\Lambda+L_{\Lambda c}\cos\theta_\Lambda+L_{\Lambda 2c} \cos2\theta_\Lambda\bigg)/\Gamma,\label{dwL}
\end{eqnarray} 
where
\begin{eqnarray} 
L_{\Lambda}&=&\mathcal{P}\frac{4\pi}{3}(3 L_{11}-L_{31}), \quad L_{\Lambda c}=\mathcal{P}\frac{4\pi}{3}(3 L_{12}-L_{32}),\quad L_{\Lambda 2c}=\mathcal{P}\frac{4\pi}{3}(3L_{13}-L_{33}).
\end{eqnarray}
The $\cos\theta_{\Lambda}$ distributions are described in Fig.~\ref{decay width}.
In the $L_{\Lambda}$ all three resonances contribute, while the $\cos2\theta_{\Lambda}$  term receives no contribution from spin-$\frac{1}{2}$ baryon and the $L_{\Lambda_{c}}$ corresponds to the interference of spin-$\frac{1}{2}$ and spin-$\frac{3}{2}$ resonance.


Based on this interference, one can construct a  normalized forward-backward asymmetry $A^\Lambda_{FB}$ of  angle $\theta_\Lambda$: 
\begin{eqnarray}
 A^\Lambda_{FB} &=&\frac{\big[\int^1_0-\int^0_{-1}\big]d\cos\theta_\Lambda\frac{d^2\Gamma}{dM_{pK}^2 d\cos\theta_\Lambda}}{\big[\int^1_0+\int^0_{-1}\big]d\cos\theta_\Lambda\frac{d^2\Gamma}{dM_{pK}^2 d\cos\theta_\Lambda}}=
\frac{3\bigg(3L_{12}-L_{32}\bigg)}{2\bigg(9L_{11}-3L_{13}-3L_{31}+L_{33}\bigg)}=\frac{3L_{\Lambda c}}{2(3L_\Lambda-L_{\Lambda2c})}. 
\end{eqnarray}

\begin{figure}[htbp!]
  \begin{minipage}[t]{0.4\linewidth}
  \includegraphics[width=1.0\columnwidth]{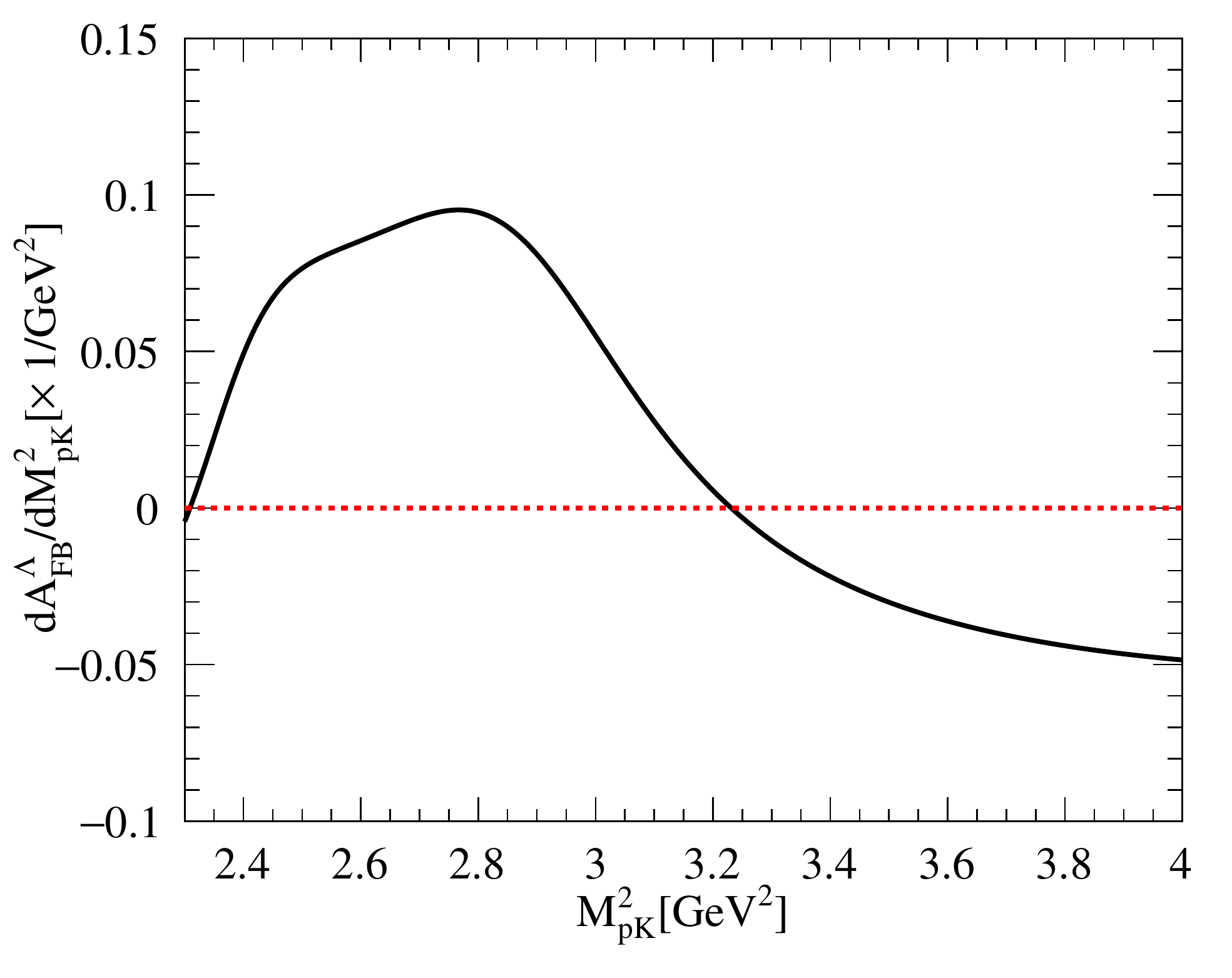}
    \end{minipage}
 \caption{The $dA^\Lambda_{FB}/dM_{pK}^2$ of process $\Lambda_b\to\Lambda_J^*(pK)J/\psi(\ell^+\ell^-)$ for $\ell=\mu$.} \label{decay widthAFB}
\end{figure}

\begin{figure}[htbp!]
  \begin{minipage}[t]{0.4\linewidth}
  \centering
  \includegraphics[width=1.0\columnwidth]{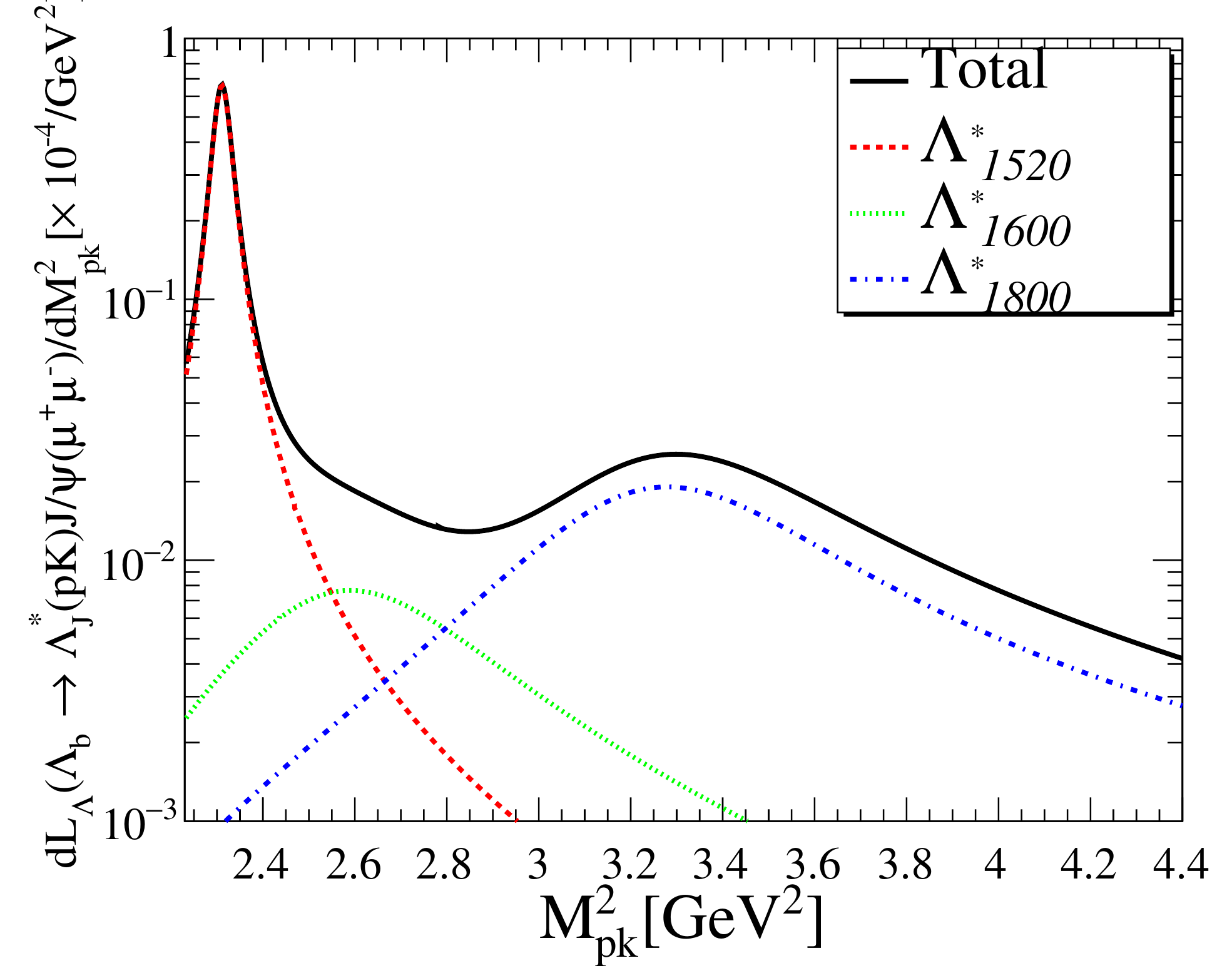}
    \end{minipage}
    \begin{minipage}[t]{0.4\linewidth}
  \centering
  \includegraphics[width=1.0\columnwidth]{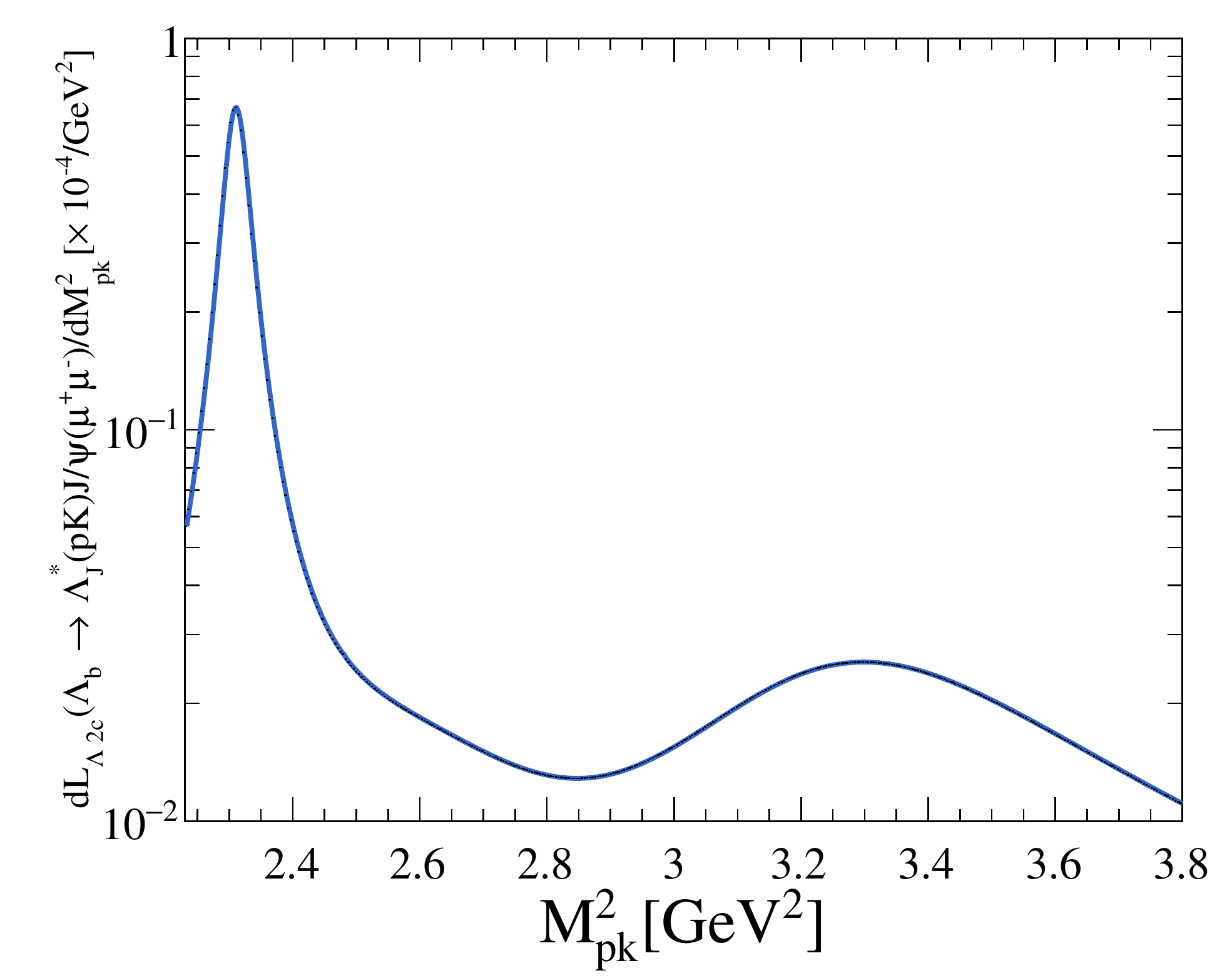}
    \end{minipage}
 \caption{The coefficients $L_{\Lambda}$ and $L_{\Lambda 2c}$  in Eq.\eqref{dwL} for $\Lambda_b\to\Lambda^*(pK)J/\psi(\ell^+\ell^-),\ell=\mu$. }\label{thLL}
 \end{figure}

Results for $A^\Lambda_{FB}$ are given in Fig.~\ref{decay widthAFB}. It is interesting to notice that the forward-backward asymmetry has a crossing point, which satisfies
\begin{eqnarray} 
\frac{dA^\Lambda_{FB}}{dM_{pK}^2}&\propto&
\frac{4\pi}{3}\bigg(3L_{12}-L_{32}\bigg)=0,\label{czp}
\end{eqnarray} 
or
\begin{eqnarray} 
&&\sum_{s_{\Lambda_{b}},s_{\Lambda_J^{*}}=\pm\frac{1}{2}}(2\hat{m}_\ell^2+1)\mathcal{R}_e(H^{\frac{3}{2}}_{s_{\Lambda_{b}},s_{\Lambda_J^{*}}}H^{\frac{1}{2}*}_{s_{\Lambda_{b}},s_{\Lambda_J^{*}}})=\notag\\
&&\sum_{s_{\Lambda_{b}},s_{\Lambda_J^{*}}=\pm\frac{1}{2}}(2\hat{m}_\ell^2+1)\big(\mathcal{R}_e(H^{\frac{3}{2}}_{s_{\Lambda_{b}},s_{\Lambda_J^{*}}})\mathcal{R}_e(H^{\frac{1}{2}*}_{s_{\Lambda_{b}},s_{\Lambda_J^{*}}})-\mathcal{I}_m(H^{\frac{3}{2}}_{s_{\Lambda_{b}},s_{\Lambda_J^{*}}})\mathcal{I}_m(H^{\frac{1}{2}*}_{s_{\Lambda_{b}},s_{\Lambda_J^{*}}})\big)=0.\label{cp}
\end{eqnarray}
It can be seen from Fig.\ref{decay widthAFB} that there are two cross point $s^1_0$ and $s^2_0$: 
\begin{eqnarray}
s_0^1=2.307\rm{GeV}^2,\quad s_0^2=3.231\rm{GeV}^2.
\end{eqnarray}
The two points are very  close to the invariant mass square of $\Lambda^*_{1520,1800}$: $m_{\Lambda_{1520}^{*}}^2=2.308\rm{GeV}^2,m_{\Lambda_{1800}^{*}}^2=3.240\rm{GeV}^2$. As shown in Fig.~\ref{decay width}, the  contribution of $\Lambda^*_{1600}$ is tiny and can be neglected. Therefore in this scenario Eq.\eqref{cp} becomes
\begin{eqnarray} 
\frac{dA^\Lambda_{FB}}{dM_{pK}^2}\propto\sum_{s_{\Lambda_{b}},s_{\Lambda_{J}^{*}}=\pm\frac{1}{2}}(2\hat{m}_\ell^2+1)\mathcal{R}_e(H^{\frac{3}{2}}_{s_{\Lambda_{b}},s_{\Lambda_{J}^{*}}}H^{\frac{1}{2}*}_{s_{\Lambda_{b}},s_{\Lambda_{J}^{*}}})\propto \mathcal{R}_e(L_{\Lambda^{*}_{1520}}L_{\Lambda^{*}_{1800}}).
\end{eqnarray} 
 The complex phase in $H^{J}_{s_{\Lambda_{b}},s_{\Lambda_J^{*}}}$ comes from the lineshape $L_{\Lambda^*_J}$, while the imaginary part is proportional to the $\Gamma_{\Lambda_J^*}m_{\Lambda_J^*}$. One can ignore the imaginary part, due to the small $\Gamma_{\Lambda_J^*}$. Thus the forward-backward asymmetry will mostly be determined by lineshape $L_{\Lambda^*_J}$ and the equation becomes
 \begin{eqnarray} 
&& \mathcal{R}_e(L_{\Lambda*_{1520}}L^*_{\Lambda*_{1800}})\sim(M_{pK}^2-m_{\Lambda^*_{1520}}^2)(M_{pK}^2-m_{\Lambda^*_{1800}}^2)=0.
\end{eqnarray} 
Thus the $s^1_0$ and $s^2_0$ should be close to the mass square of $\Lambda^*_{1520,1800}$. It will be a new method for precisely measuring resonant mass in experiments. 
Besides, one can find that the $ A^\Lambda_{FB}$ is positive in the region $M_{pK}^{2}$=$[s_0^1,s_0^2]$ and negative when $M_{pK}^{2}$ is larger than $s_0^{2}$. Therefore the two parts will almost cancel each other when  the $M_{pK}^2$ is integrated out in $ A^\Lambda_{FB}$.  The coefficient $L_{\Lambda c}$ in Eq.~\eqref{dwL} has the same behavior with $ A^\Lambda_{FB}$  and it will also give a small value. This conclusion is also confirmed by our numerical analysis for integrating $L_{\Lambda c}$ with $M_{pK}^2$ as
 \begin{eqnarray}
\int dM^2_{pK}  L_{\Lambda c}=1.95\times 10^{-5}.
 \end{eqnarray} 
Thus Fig.~3 shows the nearly symmetric curve in $\cos\theta_{\Lambda}$ distribution.
Besides, we show the  results for $(L_{\Lambda}, L_{\Lambda 2c})$ distributions in Fig.~\ref{thLL}.
It can be seen that only the spin-$\frac{3}{2}$ resonance  contributes to the coefficient $L_{\Lambda2c}$, and thus this angular coefficient gives a piece of clear information on the spin-$\frac{3}{2}$ resonance. 

\subsubsection{Distribution in the azimuthal angle $\phi$} 

The normalized angular distribution in $\phi$ can be derived by integrating  the angle $(\theta_\Lambda,\theta)$,
\begin{eqnarray} 
\frac{1}{\Gamma}\frac{d^2\Gamma(\Lambda_b\to\Lambda_J^*(pK)J/\psi(\ell^+\ell^-))}{ dM_{pK}^2d\phi}&=&\bigg(L_\phi+L_{\phi 2c}\cos2\phi+L_{\phi 2s} \sin2\phi\bigg)/\Gamma,\label{phi}
\end{eqnarray} 
where
\begin{eqnarray}
&&L_{\phi}=\mathcal{P}\frac{4}{9}(9 L_{11}-(3L_{31}+3L_{13})+L_{33}),\notag\\
&& L_{\phi 2c}=\mathcal{P}\frac{4}{9}(9 L_{21}-(3L_{22}+3L_{51})+L_{52}),\notag\\
&& L_{\phi 2s}=\mathcal{P}\frac{4}{9}(9 L_{71}-(3L_{72}+3L_{81})+L_{82}).
\end{eqnarray}
For these three coefficients, the numerical results $(L_{\phi},L_{\phi 2c},L_{\phi 2s})$ are given in Fig.~\ref{phi}. One can see that in Eq.~\eqref{coe} only the interference of different polarisation helicity amplitudes of $\Lambda_{1520}^*$ can contribute to $L_{\phi 2s}$.  Since the complex phase in the helicity amplitude comes from the Breit-Wigner lineshape, the coefficients $L_{71}, L_{72}, L_{81}$ and $L_{82}$ are equal to zero. Therefore the coefficient $L_{\phi 2s}$ is vanishing.
\begin{figure}[htbp!]
  \begin{minipage}[t]{0.4\linewidth}
  \centering
  \includegraphics[width=1.0\columnwidth]{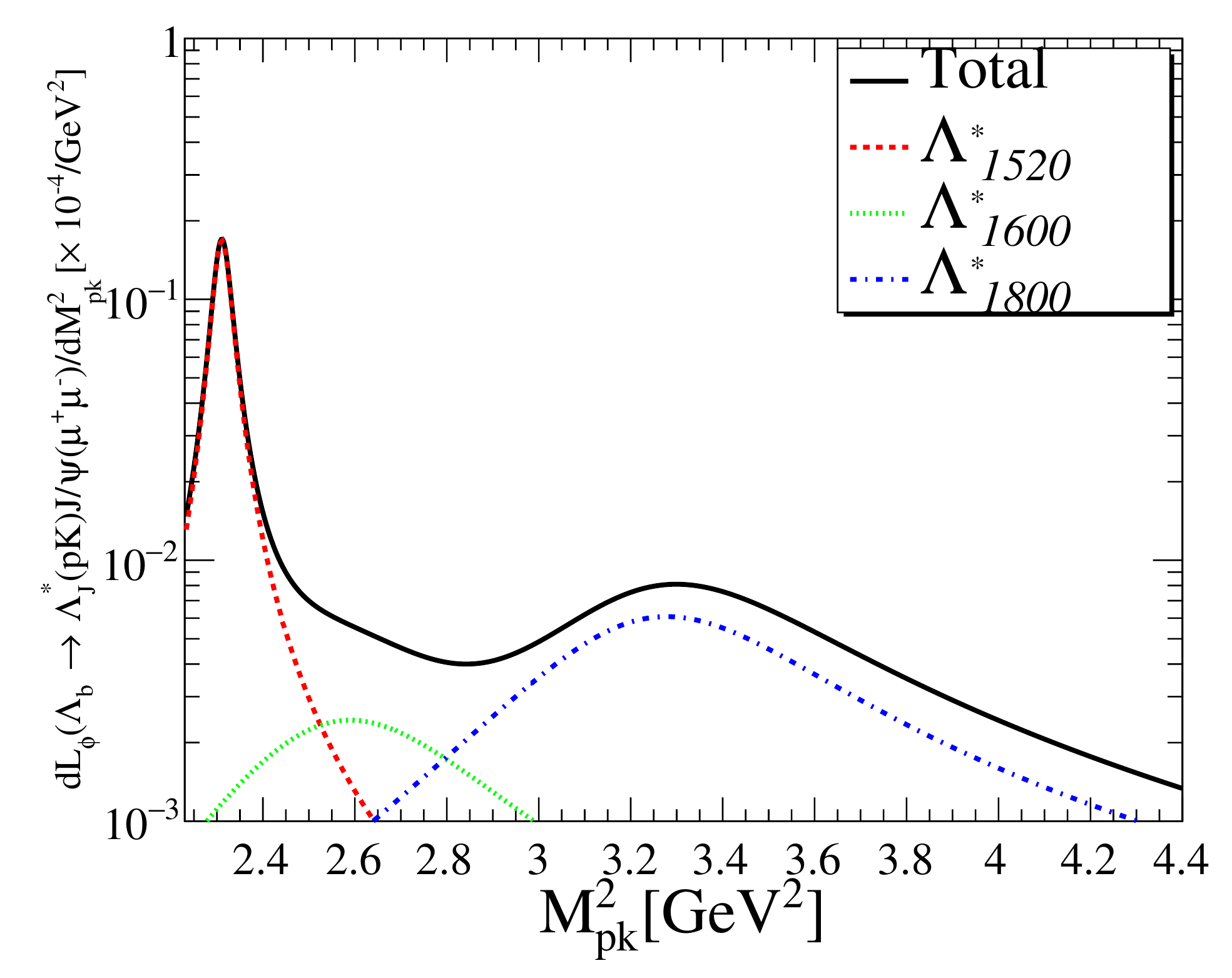}
    \end{minipage}
    \begin{minipage}[t]{0.4\linewidth}
  \centering
  \includegraphics[width=1.0\columnwidth]{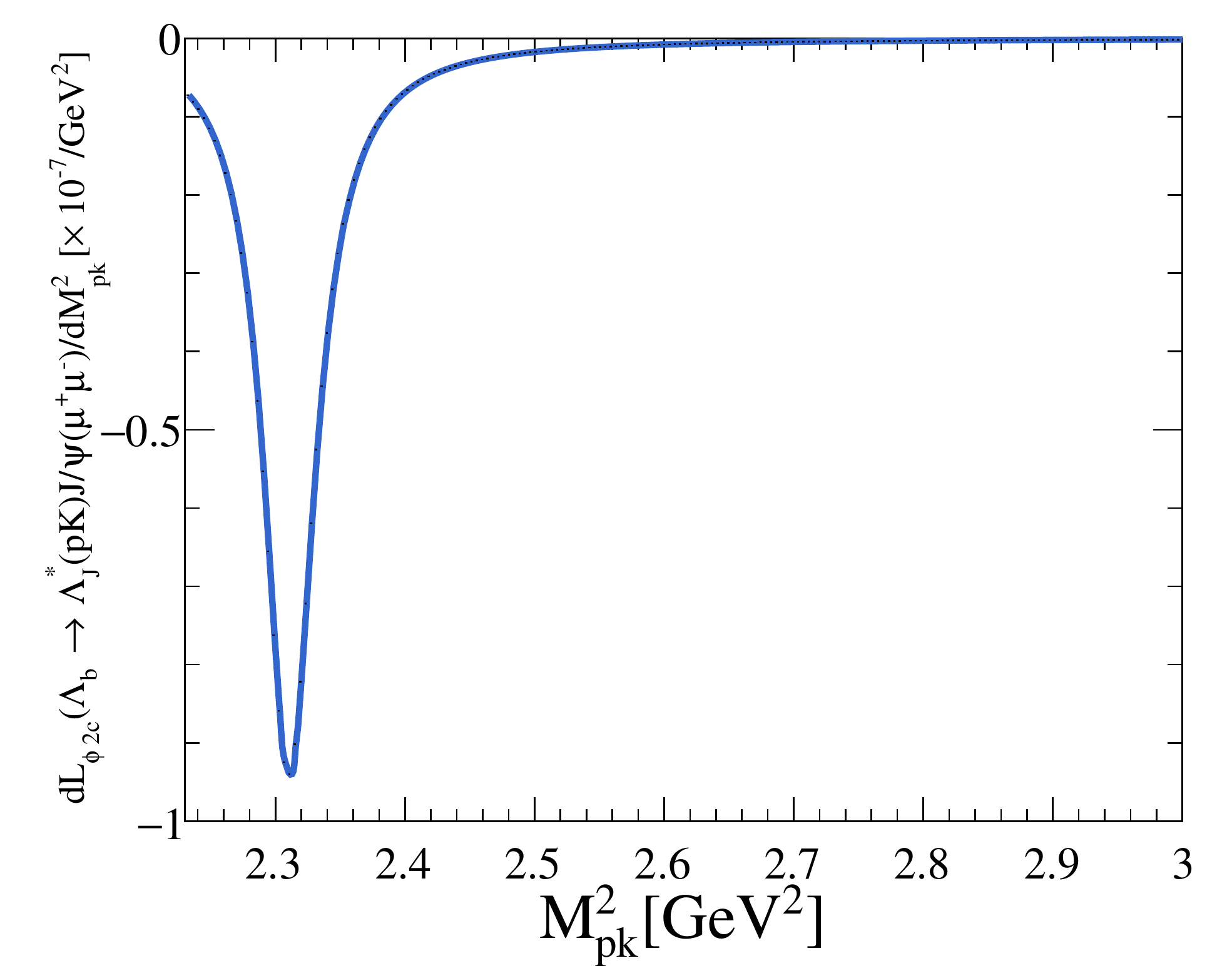}
    \end{minipage}
 \caption{The coefficients $L_{\phi}$ and $L_{2c\phi}$  in Eq.\eqref{phi} for $\Lambda_b\to\Lambda_J^*(pK)J/\psi(\ell^+\ell^-),\ell=\mu,e$.  }\label{phi}
 \end{figure}

 We can see that $L_{\phi}$ has the same behavior with Eq.\eqref{dw} and the numerical results of $L_{\phi 2c}$ are tiny shown in Fig.\ref{phi}. It is due to the $\mathcal{R}_e(H^{\frac{3}{2}}_{\frac{1}{2},\frac{3}{2}}H^{\frac{3}{2}*}_{\frac{1}{2},-\frac{1}{2}})$ term in the coefficient $L_{21},L_{22},L_{51}$ and $L_{52}$ are cancelled with each other.

\subsubsection{Polarisation of the  $\Lambda_b$}

\begin{figure}[htbp!]
  \centering
  \includegraphics[width=0.6\columnwidth]{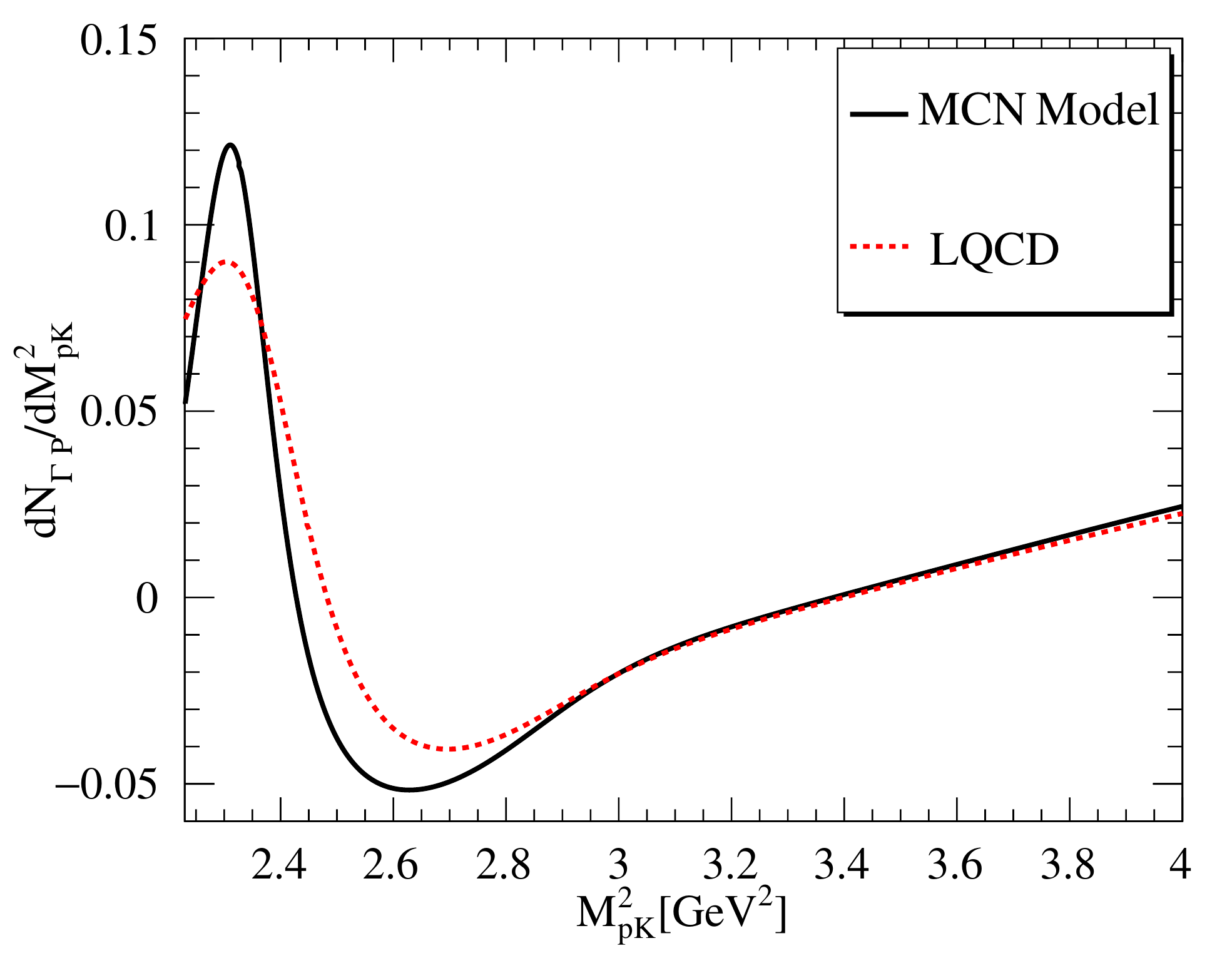}
 \caption{The normalized polarised decay width $dN_{\Gamma_P}/dM^2_{pK}$ of  $\Lambda_b(s_{\Lambda_b})\to\Lambda^*_J(pK)J/\psi(\mu^+\mu^-)$. The black solid line is utilized the MCN quark model form factors and the red doted one is drawn by LQCD form factors for resonance $\Lambda^*_{1520}$ and the MCN quark model form factors for $\Lambda^*_{1600,1800}$.}\label{plb}
 \end{figure}

The polarised angular distribution of $s_{\Lambda_b}$ can be described as
\begin{eqnarray}
\frac{d\Gamma(s_{\Lambda_b})}{d\cos\theta d\cos\theta_\Lambda d\phi dM_{pK}^2}&=&\mathcal{P}\bigg(L^{(s_{\Lambda_b})}_{11}+\cos\theta_\Lambda L^{(s_{\Lambda_b})}_{12}+\cos2\theta_\Lambda L^{(s_{\Lambda_b})}_{13}+\cos2\phi(L^{(s_{\Lambda_b})}_{21}+\cos2\theta_\Lambda L^{(s_{\Lambda_b})}_{22})+\notag\\
&&\cos2\theta(L^{(s_{\Lambda_b})}_{31}+\cos\theta_\Lambda L^{(s_{\Lambda_b})}_{32}+\cos2\theta_\Lambda L^{(s_{\Lambda_b})}_{33})+ \sin2\theta\cos\phi(\sin\theta_\Lambda L^{(s_{\Lambda_b})}_{41}+\sin2\theta_\Lambda L^{(s_{\Lambda_b})}_{42})+\notag\\
&&\cos2\phi\cos2\theta(L^{(s_{\Lambda_b})}_{51}+\cos2\theta_\Lambda L^{(s_{\Lambda_b})}_{52})+\sin2\theta\sin\phi(\sin\theta_\Lambda L^{(s_{\Lambda_b})}_{61}+\sin2\theta_{\Lambda} L^{(s_{\Lambda_b})}_{62})+\notag\\
&&\sin2\phi(L^{(s_{\Lambda_b})}_{71}+\cos2\theta_\Lambda L^{(s_{\Lambda_b})}_{72})+\cos2\theta\sin2\phi(L^{(s_{\Lambda_b})}_{81}+\cos2\theta_\Lambda L^{(s_{b})}_{82})\bigg).
\end{eqnarray}
Using the distribution polarised, the normalized polarised decay width can be defined as 
\begin{eqnarray}
\frac{dN_{\Gamma_P}}{dM^2_{pK}}=\frac{\frac{d\Gamma(\frac{1}{2})}{dM^2_{pK}}-\frac{d\Gamma(-\frac{1}{2})}{dM^2_{pK}}}{\frac{d\Gamma(\frac{1}{2})}{dM^2_{pK}}+\frac{d\Gamma(-\frac{1}{2})}{dM^2_{pK}}},\label{nP}
\end{eqnarray}
and it is shown for LQCD form factor and MCN quark model in Fig.~\ref{plb}.

The distribution of normalized polarized decay width shows a discrepancy with different polarized $\Lambda_{b}$.  The distributions of normalized polarized branching fractions with two sets of form factors are shown in Fig.\ref{plb}, which indicates the distribution of the two types of methods are similar except in the low $M_{pK}^2$ region. After normalizing the polarized decay width, the  difference caused by LQCD and MCN form factors is less significant. It is due to the fact that in the normalized  decay width, many common factors  have been canceled.  Besides, one can also find that the branching fraction with $s_{\Lambda_b}=1/2$ is larger than that with  $s_{\Lambda_b}=-1/2$ in the low-$M_{pK}^2$ region for both two sets of form factor results.  One can see that the decay width in Eq.\eqref{dwa} shows the symmetry of transformation $\Big((s_{\Lambda_b} ,s_{\Lambda_J^*})\to (-s_{\Lambda_b},-s_{\Lambda_J^*})\Big)$. Since the helicity amplitudes in the appendix~\ref{apA} show that for vector current and axis vector current the transformation brings positive and negative signs respectively, the polarized decay width is mainly contributed by the interference of vector current and axis vector current hadron helicity amplitudes.  It is noteworthy that the interference terms between  vector current and axis vector current helicity amplitudes have no contributions to the non-polarised decay width. Therefore the polarized decay width is a very important and unique observable for the study of the hadron matrix element structure.

\section{Conclusions}

In this work, the differential and integrated decay width for the process of $\Lambda_b\to \Lambda^*_J(pK)J/\psi(\ell^+\ell^-)$ through different resonances $\Lambda^*(\frac{1}{2}^+,\frac{1}{2}^-,\frac{3}{2}^-)$ are studied. Branching fractions with the individual resonances and total results are given in  Eq.\eqref{Br} by taking into form factors of $\Lambda_b\to \Lambda^*_{1520}$ in Lattice QCD and form factors of $\Lambda_b\to \Lambda^*_{1600,1800}$ in the MCN quark model.

For this process, we have derived the angular distribution with the possible resonance $\Lambda^*_{1520,1600,1800}$ and other phenomenological results such as partial decay width, polarisation and forward-backward asymmetry with final states as muon and electron respectively. Our results with different lepton $e$ and $\mu$ are highly consistent with the lepton flavor universal. It has a good reference value for lepton flavor universal experiments. For the resonance $\Lambda^*_{1520}$, we adopt different types of form factors: the lattice QCD and MCN quark model, which shows a big discrepancy in the low$-M_{pK}^2$ region in Fig.\ref{LAQM}. Since the Lattice QCD has more reasonable results only in the high$-M_{pK}^2$ region, we give the branching fraction and differential branching fractions for both two sets of form factors.  We have analyzed the distribution of the angle and show the dependence of $M_{pK}^2$. 

Results in this work will serve as a calibration for the study of  $b\to s \bar{\mu}\mu$ decays in $\Lambda_b$ decays in future and provide useful information towards the understanding of the properties of the $\Lambda_J^*$ baryons. Recently, the LHCb Collaboration has analyzed the process $\Lambda_b\to pK^-\ell^+\ell^-$~\cite{LHCb:2019efc}. Therefore, the analysis of  the angular distribution of $\Lambda_{b}\rightarrow{\Lambda^*_J({p K})~J/\psi({\ell^{+}\ell^{-}})}$ in the LHCb is feasible and we urge our experimental colleagues to analyze this very interesting process.

\section*{Acknowledgements}
We thank Prof. Jibo He, Mr. Zhiyu Xiang and Dr. Yixiong Zhou for fruitful discussions on the potential to measure this decay channel at LHCb. 
This work  is supported in part by Natural Science Foundation of China under grant No. 11735010, 11911530088, 12147147, by Natural Science Foundation of Shanghai under grant No. 15DZ2272100.

\begin{appendix}
\section{Helicity Amplitude}\label{apA}
The hadronic helicity amplitudes we used are defined with the hadron matrix element as
\begin{eqnarray}
H_{A}(s_1,s_2,s_W)&=&
\langle \Lambda^*_J(p^\prime, s_2)|\bar s\gamma^\mu \gamma_5 b|\Lambda_b(p, s_1)\rangle\epsilon^*_\mu(q,s_W),\notag\\
H_{V}(s_1,s_2,s_W)&=&
\langle \Lambda^*_J(p^\prime, s_2)|\bar s\gamma^\mu  b|\Lambda_b(p, s_1)\rangle\epsilon^*_\mu(q,s_W).
\end{eqnarray}
We give the  hadronic helicity amplitudes for the $\Lambda_b\to \Lambda^*_{1520}$ transition: 
 \begin{align}
H_{V}\left(s_{1}=\frac{1}{2}, s_2=\frac{3}{2}, s_{W}=1\right) &=H_{V}\left(s_{1}=-\frac{1}{2}, s_2=-\frac{3}{2}, s_{W}=-1\right) \nonumber\\ 
&=-f_{4}(M_{pk}^{2}) \sqrt{s_{p+}}, \\
H_{V}\left(s_{1}=\frac{1}{2}, s_2=-\frac{1}{2}, s_{W}=-1\right) &=H_{V}\left(s_{1}=-\frac{1}{2}, s_2=\frac{1}{2}, s_{W}=1\right) \nonumber\\
&=\sqrt{\frac{s_{p+}}{3}}\left( \frac{s_{p-}}{m_{\Lambda}m_{\Lambda_{1520}^{*}}}f_{1}(M_{pk}^{2})-f_{4}(M_{pk}^{2})\right), \\
H_{V}\left(s_{1}=\frac{1}{2}, s_2=\frac{1}{2}, s_{W}=0\right)&=H_{V}\left(s_{1}=-\frac{1}{2}, s_2=-\frac{1}{2}, s_{W}=0\right) \nonumber\\
&= \sqrt{\frac{s_{p+}}{6}}\frac{1}{\mjp}\left[ s_{p-}s_{p+}\left(\frac{f_{1}(M_{Pk}^{2})}{s_{p+}}(\frac{1}{m_{\Lambda_{1520}^{*}}}+\frac{1}{\mI})\right.\right.\nonumber\\
&\left.\quad +f_{2}(M_{Pk}^{2})\frac{1}{2m_{\Lambda_{1520}^{*}}\mI^2}+f_{3}(M_{Pk}^{2})\frac{1}{2m_{\Lambda_{1520}^{*}}\mI\sqrt{M_{pk}^2}}\right)\nonumber\\
&\left.\quad +f_{4}(M_{Pk}^{2})\frac{\mI^2-\mjp^2-M_{pk}^2}{m_{\Lambda_{1520}^{*}}}\right], \\
H_{A}\left(s_{1}=\frac{1}{2}, s_2=\frac{3}{2}, s_{W}=1\right) &=-H_{A}\left(s_{1}=-\frac{1}{2}, s_2=-\frac{3}{2}, s_{W}=-1\right) \nonumber\\ 
&=g_{4}(M_{Pk}^{2})\sqrt{s_{p-}},  \\
H_{A}\left(s_{1}=\frac{1}{2}, s_2=-\frac{1}{2}, s_{W}=-1\right) &=-H_{A}\left(s_{1}=-\frac{1}{2}, s_2=\frac{1}{2}, s_{W}=1\right) \nonumber\\
&=\sqrt{\frac{s_{p-}}{3}}\left(\frac{s_{p+}}{m_{\Lambda_{1520}^{*}}\mI}g_{1}(M_{Pk}^{2})-g_{4}(M_{Pk}^{2})\right), \\
H_{A}\left(s_{1}=\frac{1}{2}, s_2=\frac{1}{2}, s_{W}=0\right)&=-H_{A}\left(s_{1}=-\frac{1}{2}, s_2=-\frac{1}{2}, s_{W}=0\right) \nonumber\\
&=\sqrt{\frac{s_{p-}}{6}}\frac{1}{\mjp}\left[ s_{p-}s_{p+}\left(\frac{g_{1}(M_{Pk}^{2})}{s_{p-}}(\frac{1}{m_{\Lambda_{1520}^{*}}}-\frac{1}{\mI})\right.\right.\nonumber\\
&\left.\quad -g_{2}(M_{Pk}^{2})\frac{1}{2m_{\Lambda_{1520}^{*}}\mI^2}-g_{3}(M_{Pk}^{2})\frac{1}{2m_{\Lambda_{1520}^{*}}\mI\sqrt{M_{pk}^2}}\right)\nonumber\\
&\left.\quad +g_{4}(M_{Pk}^{2})\frac{-\mI^2+\mjp^2+M_{pk}^2}{m_{\Lambda_{1520}^{*}}}\right].
\end{align}


If the spin of the final state is one half, the helicity amplitude is given as:
 \begin{align}
H_{ V}\left(s_1=-\frac{1}{2}, s_2=\frac{1}{2}, s_W=1\right) &=H_{V}\left(s_1=\frac{1}{2}, s_2=-\frac{1}{2}, s_W=-1\right) \nonumber\\  
&=\sqrt{2 }\left[f_{1}(M_{pK}^2)\sqrt{s_{p-}}\right], \\
H_{ V}\left(s_1=\frac{1}{2}, s_2=\frac{1}{2}, s_W=0\right) &=H_{V}\left(s_1=-\frac{1}{2}, s_2=-\frac{1}{2}, s_W=0\right) \nonumber\\
&=\frac{\sqrt{s_{p-}}}{m_{J/\psi}}\left[\left(\mI+\mf\right) f_{1}(M_{pK}^{2})+\frac{s_{p+}}{2m_{\Lambda_{b}}} f_{2}(M_{pK}^{2})+f_{3}\frac{s_{p+}}{2\sqrt{M_{pK}^{2}}}\right],\nonumber\\
H_{A}\left(s_1=-\frac{1}{2}, s_2=\frac{1}{2}, s_W=1\right) &=-H_{ A}\left(s_1=\frac{1}{2}, s_2=-\frac{1}{2}, s_W=-1\right) \nonumber\\
&=\sqrt{2 s_{p+}}\left[g_{1}(M_{pK}^{2}))\right], \\
H_{ A}\left(s_1=\frac{1}{2}, s_2=\frac{1}{2}, s_W=0\right) &=-H_{ A}\left(s_1=-\frac{1}{2}, s_2=-\frac{1}{2}, s_W=0\right) \nonumber\\
&=\frac{\sqrt{s_{p+}}}{m_{J/\psi}}\left[\left(-\mI+\mf\right) g_{1}(M_{pK}^{2})+\frac{s_{p-}}{2m_{\Lambda_{b}}} g_{2}(M_{pK}^{2})+g_{3}\frac{s_{p-}}{2\sqrt{M_{pK}^{2}}}\right].
\end{align}

The leptonic helicity amplitudes $L_{s_-,s_+}^{s_{J/\psi}}$ are
 \begin{eqnarray}
 &&L^1_{\frac{1}{2},\frac{1}{2}}(\phi,\theta)=L^1_{-\frac{1}{2},-\frac{1}{2}}(\phi,\theta)=-i\sqrt{2}m_\ell e^{-i\phi}\sin\theta,\;L^1_{\frac{1}{2},-\frac{1}{2}}(\phi,\theta)=-i\frac{m_{J/\psi}}{\sqrt{2}}m_\ell e^{-i\phi}(\cos\theta+1),\notag\\
 &&L^1_{-\frac{1}{2},\frac{1}{2}}(\phi,\theta)=i\frac{m_{J/\psi}}{\sqrt{2}}m_\ell e^{-i\phi}(\cos\theta-1),\quad\quad\quad L^0_{\frac{1}{2},\frac{1}{2}}(\phi,\theta)=L^0_{-\frac{1}{2},-\frac{1}{2}}(\phi,\theta)=-2im_\ell\cos\theta,\notag\\
  &&L^{-1}_{-\frac{1}{2},-\frac{1}{2}}(\phi,\theta)=L^{-1}_{-\frac{1}{2},-\frac{1}{2}}(\phi,\theta)=i\sqrt{2}m_\ell e^{i\phi}\sin\theta,\;L^{-1}_{-\frac{1}{2},\frac{1}{2}}(\phi,\theta)=-i\frac{m_{J/\psi}}{\sqrt{2}}m_\ell e^{i\phi}(\cos\theta+1),\notag\\
 &&L^{-1}_{\frac{1}{2},-\frac{1}{2}}(\phi,\theta)=i\frac{m_{J/\psi}}{\sqrt{2}}m_\ell e^{i\phi}(\cos\theta-1).
\end{eqnarray}
For the $J=\frac{1}{2}$ resonances $\Lambda^*_J$, the Wigner functions are
 \begin{eqnarray}
 D^{\frac{1}{2}}_{\frac{1}{2},\frac{1}{2}}(\phi_\Lambda,\theta_\Lambda)&=&e^{-i\frac{1}{2}\phi_\Lambda}\cos(\frac{1}{2}\theta_\Lambda),\quad
  D^{\frac{1}{2}}_{\frac{1}{2},-\frac{1}{2}}(\phi_\Lambda,\theta_\Lambda)=-e^{-i\frac{1}{2}\phi_\Lambda}\sin(\frac{1}{2}\theta_\Lambda),\notag\\
   D^{\frac{1}{2}}_{-\frac{1}{2},\frac{1}{2}}(\phi_\Lambda,\theta_\Lambda)&=&e^{-i\frac{1}{2}\phi_\Lambda}\sin(\frac{1}{2}\theta_\Lambda),\quad
    D^{\frac{1}{2}}_{-\frac{1}{2},-\frac{1}{2}}(\phi_\Lambda,\theta_\Lambda)=e^{-i\frac{1}{2}\phi_\Lambda}\cos(\frac{1}{2}\theta_\Lambda).
 \end{eqnarray}
 For the $J=\frac{3}{2}$ resonances $\Lambda^*_J$, the Wigner functions are
 \begin{eqnarray}
  &&D^{\frac{1}{2}}_{\frac{3}{2},\frac{1}{2}}(\phi_\Lambda,\theta_\Lambda)=-\sqrt{3}e^{-i\frac{3}{2}\phi_\Lambda}\frac{1+\cos\theta_\Lambda}{2}\sin(\frac{1}{2}\theta_\Lambda),\quad
  D^{\frac{1}{2}}_{\frac{3}{2},-\frac{1}{2}}(\phi_\Lambda,\theta_\Lambda)=\sqrt{3}e^{-i\frac{3}{2}\phi_\Lambda}\frac{1-\cos\theta_\Lambda}{2}\cos(\frac{1}{2}\theta_\Lambda),\notag\\
  &&D^{\frac{1}{2}}_{\frac{1}{2},\frac{1}{2}}(\phi_\Lambda,\theta_\Lambda)=e^{-i\frac{1}{2}\phi_\Lambda}\frac{3\cos\theta_\Lambda-1}{2}\cos(\frac{1}{2}\theta_\Lambda),\quad D^{\frac{1}{2}}_{\frac{1}{2},-\frac{1}{2}}(\phi_\Lambda,\theta_\Lambda)=-e^{-i\frac{1}{2}\phi_\Lambda}\frac{3\cos\theta_\Lambda+1}{2}\sin(\frac{1}{2}\theta_\Lambda),\notag\\
   &&D^{\frac{1}{2}}_{-\frac{3}{2},-\frac{1}{2}}(\phi_\Lambda,\theta_\Lambda)=\sqrt{3}e^{i\frac{3}{2}\phi_\Lambda}\frac{1+\cos\theta_\Lambda}{2}\sin(\frac{1}{2}\theta_\Lambda),\quad
  D^{\frac{1}{2}}_{\frac{3}{2},-\frac{1}{2}}(\phi_\Lambda,\theta_\Lambda)=\sqrt{3}e^{i\frac{3}{2}\phi_\Lambda}\frac{1-\cos\theta_\Lambda}{2}\cos(\frac{1}{2}\theta_\Lambda),\notag\\
   &&D^{\frac{1}{2}}_{-\frac{1}{2},-\frac{1}{2}}(\phi_\Lambda,\theta_\Lambda)=e^{i\frac{1}{2}\phi_\Lambda}\frac{3\cos\theta_\Lambda-1}{2}\cos(\frac{1}{2}\theta_\Lambda),\quad D^{\frac{1}{2}}_{-\frac{1}{2},\frac{1}{2}}(\phi_\Lambda,\theta_\Lambda)=e^{i\frac{1}{2}\phi_\Lambda}\frac{3\cos\theta_\Lambda+1}{2}\sin(\frac{1}{2}\theta_\Lambda).
  \end{eqnarray}
  
  \section{Coefficient function in angular distribution}\label{coef}
  
  The specific expressions of coefficient $L_i$ in containing the resonance $\Lambda^*_{1520,1600,1800}$ are
  \begin{eqnarray}
L_1&=&\sum_{s_p}\bigg((4\hat{m}_\ell^2+1)\Big(\bigg| H^{\frac{1}{2}}_{\frac{1}{2},\frac{1}{2}}D^{\frac{1}{2}}_{\frac{1}{2},s_p}(0,\theta_\Lambda) \bigg| ^2+\bigg| H^{\frac{3}{2}}_{\frac{1}{2},\frac{1}{2}}D^{\frac{3}{2}}_{\frac{1}{2},s_p}(0,\theta_\Lambda) \bigg| ^2+2\mathcal{R}_e(H^{\frac{1}{2}}_{\frac{1}{2},\frac{1}{2}}H^{\frac{3}{2}*}_{\frac{1}{2},\frac{1}{2}})D^{\frac{1}{2}*}_{\frac{1}{2},s_p}(0,\theta_\Lambda)D^{\frac{3}{2}}_{\frac{1}{2},s_p}(0,\theta_\Lambda)\Big)\notag\\
&&+\frac{1}{2}(4\hat{m}_\ell^2+3)\Big(\bigg| H^{\frac{1}{2}}_{\frac{1}{2},-\frac{1}{2}}D^{\frac{1}{2}}_{-\frac{1}{2},s_p}(0,\theta_\Lambda) \bigg| ^2+\bigg| H^{\frac{3}{2}}_{\frac{1}{2},-\frac{1}{2}}D^{\frac{3}{2}}_{-\frac{1}{2},s_p}(0,\theta_\Lambda) \bigg| ^2+\bigg| H^{\frac{3}{2}}_{\frac{1}{2},\frac{3}{2}}D^{\frac{3}{2}}_{\frac{3}{2},s_p}(0,\theta_\Lambda) \bigg| ^2\notag\\
&&+2\mathcal{R}_e(H^{\frac{1}{2}}_{\frac{1}{2},-\frac{1}{2}}H^{\frac{3}{2}*}_{\frac{1}{2},-\frac{1}{2}})D^{\frac{1}{2}*}_{-\frac{1}{2},s_p}(0,\theta_\Lambda)D^{\frac{3}{2}}_{-\frac{1}{2},s_p}(0,\theta_\Lambda)\Big)\bigg)+\Big((s_{\Lambda_b} ,s_{\Lambda_{J}^*})\to (-s_{\Lambda_b},-s_{\Lambda_J^*})\Big),\notag\\
L_2&=&-(4\hat{m}_\ell^2-1)\sum_{s_p}\sum_{J=\frac{1}{2},\frac{3}{2}}\mathcal{R}_e(H^{J}_{\frac{1}{2},-\frac{1}{2}}H^{\frac{3}{2}*}_{\frac{1}{2},\frac{3}{2}})D^{*J}_{-\frac{1}{2},s_p}(0,\theta_\Lambda)D^{\frac{3}{2}}_{\frac{3}{2},s_p}(0,\theta_\Lambda)+\Big((s_{\Lambda_b} ,s_{\Lambda_J^*})\to (-s_{\Lambda_b},-s_{\Lambda_J^*})\Big),\notag\\
L_3&=&\frac{-1}{2}(4\hat{m}_\ell^2-1)\sum_{s_p}\bigg( \bigg| H^{\frac{1}{2}}_{\frac{1}{2},-\frac{1}{2}}D^{\frac{1}{2}}_{-\frac{1}{2},s_p}(0,\theta_\Lambda) \bigg| ^2+ \bigg| H^{\frac{3}{2}}_{\frac{1}{2},-\frac{1}{2}}D^{\frac{3}{2}}_{-\frac{1}{2},s_p}(0,\theta_\Lambda) \bigg| ^2\notag\\
&&+2\mathcal{R}_e(H^{\frac{1}{2}}_{\frac{1}{2},-\frac{1}{2}}H^{\frac{3}{2}*}_{\frac{1}{2},-\frac{1}{2}})D^{\frac{1}{2}*}_{-\frac{1}{2},s_p}(0,\theta_\Lambda)D^{\frac{3}{2}}_{-\frac{1}{2},s_p}(0,\theta_\Lambda)-4\mathcal{R}_e(H^{\frac{1}{2}}_{\frac{1}{2},\frac{1}{2}}H^{\frac{3}{2}*}_{\frac{1}{2},\frac{1}{2}})D^{\frac{1}{2}*}_{\frac{1}{2},s_p}(0,\theta_\Lambda)D^{\frac{3}{2}}_{\frac{1}{2},s_p}(0,\theta_\Lambda),\notag\\
&&+\bigg| H^{\frac{3}{2}}_{\frac{1}{2},\frac{3}{2}}D^{\frac{3}{2}}_{\frac{3}{2},s_p}(0,\theta_\Lambda) \bigg| ^2-2\bigg| H^{\frac{1}{2}}_{\frac{1}{2},\frac{1}{2}}D^{\frac{1}{2}}_{\frac{1}{2},s_p}(0,\theta_\Lambda) \bigg| ^2-2\bigg| H^{\frac{3}{2}}_{\frac{1}{2},\frac{1}{2}}D^{\frac{3}{2}}_{\frac{1}{2},s_p}(0,\theta_\Lambda) \bigg| ^2\bigg)+\Big((s_{\Lambda_b} ,s_{\Lambda_J^*})\to (-s_{\Lambda_b},-s_{\Lambda_J^*})\Big),\notag\\
L_4&=&-\sqrt{2}(4\hat{m}_\ell^2-1)\sum_{s_p}\sum_{J,J^\prime=\frac{1}{2},\frac{3}{2}}\bigg(\mathcal{R}_e(H^{J}_{\frac{1}{2},\frac{1}{2}}H^{*J^\prime}_{\frac{1}{2},-\frac{1}{2}})D^{*J}_{\frac{1}{2},s_p}(0,\theta_\Lambda)D^{J^\prime}_{-\frac{1}{2},s_p}(0,\theta_\Lambda),\notag\\
&&-\mathcal{R}_e(H^{J}_{\frac{1}{2},\frac{1}{2}}H^{\frac{3}{2}*}_{\frac{1}{2},\frac{3}{2}})D^{*J}_{\frac{1}{2},s_p}(0,\theta_\Lambda)D^{\frac{3}{2}}_{\frac{3}{2},s_p}(0,\theta_\Lambda)\bigg)-\Big((s_{\Lambda_b} ,s_{\Lambda_J^*})\to (-s_{\Lambda_b},-s_{\Lambda_J^*})\Big),\notag\\
L_5&=&-L_2,\notag\\
L_6&=&-\sqrt{2}(4\hat{m}_\ell^2-1)\sum_{s_p}\sum_{J,J^\prime=\frac{1}{2},\frac{3}{2}}\bigg(\mathcal{I}_m(H^{J}_{\frac{1}{2},\frac{1}{2}}H^{*J^\prime}_{\frac{1}{2},-\frac{1}{2}})D^{*J}_{\frac{1}{2},s_p}(0,\theta_\Lambda)D^{J^\prime}_{-\frac{1}{2},s_p}(0,\theta_\Lambda)\notag\\
&&+\mathcal{I}_m(H^{J}_{\frac{1}{2},\frac{1}{2}}H^{\frac{3}{2}*}_{\frac{1}{2},\frac{3}{2}})D^{*J}_{\frac{1}{2},s_p}(0,\theta_\Lambda)D^{\frac{3}{2}}_{\frac{3}{2},s_p}(0,\theta_\Lambda)\bigg)+\Big((s_{\Lambda_b} ,s_{\Lambda_J^*})\to (-s_{\Lambda_b},-s_{\Lambda_J^*})\Big),\notag\\
L_7&=&(4\hat{m}_\ell^2-1)\sum_{s_p}\sum_{J=\frac{1}{2},\frac{3}{2}}\mathcal{I}_m(H^{J}_{\frac{1}{2},-\frac{1}{2}}H^{\frac{3}{2}*}_{\frac{1}{2},\frac{3}{2}})D^{*J}_{-\frac{1}{2},s_p}(0,\theta_\Lambda)D^{\frac{3}{2}}_{\frac{3}{2},s_p}(0,\theta_\Lambda)-\Big((s_{\Lambda_b} ,s_{\Lambda_J^*})\to (-s_{\Lambda_b},-s_{\Lambda_J^*})\Big),\notag\\
L_8&=&-L_7.
\end{eqnarray}

Here both $J$ and $J^\prime$ represent the spin of resonant state $\Lambda^{*}_{J}$.

The formulas of coefficient function $L_{ij}(i=1-8,j=1-3)$ are given as
\begin{eqnarray}
L_{11}&=&\frac{1}{4}\hat{m}_\ell^2\big(3|H^{\frac{3}{2}}_{\frac{1}{2},\frac{3}{2}}|^2+16|H^{\frac{1}{2}}_{\frac{1}{2},\frac{1}{2}}|^2+10|H^{\frac{3}{2}}_{\frac{1}{2},\frac{1}{2}}|^2+5|H^{\frac{3}{2}}_{\frac{1}{2},-\frac{1}{2}}|^2+8|H^{\frac{1}{2}}_{\frac{1}{2},-\frac{1}{2}}|^2\big)+\notag\\
&&\frac{1}{16}\big(9|H^{\frac{3}{2}}_{\frac{1}{2},\frac{3}{2}}|^2+16|H^{\frac{1}{2}}_{\frac{1}{2},\frac{1}{2}}|^2+10|H^{\frac{3}{2}}_{\frac{1}{2},\frac{1}{2}}|^2+15|H^{\frac{3}{2}}_{\frac{1}{2},-\frac{1}{2}}|^2+24|H^{\frac{1}{2}}_{\frac{1}{2},-\frac{1}{2}}|^2\big)+\Big((s_{\Lambda_b} ,s_{\Lambda_J^*})\to (-s_{\Lambda_b},-s_{\Lambda_J^*})\Big),\notag\\
L_{12}&=&2(4\hat{m}_\ell^2+1)\mathcal{R}_e(H^{\frac{3}{2}}_{\frac{1}{2},\frac{1}{2}}H^{\frac{1}{2}*}_{\frac{1}{2},\frac{1}{2}})+(4\hat{m}_\ell^2+3)\mathcal{R}_e(H^{\frac{3}{2}}_{-\frac{1}{2},\frac{1}{2}}H^{\frac{1}{2}*}_{-\frac{1}{2},\frac{1}{2}})+\Big((s_{\Lambda_b} ,s_{\Lambda_J^*})\to (-s_{\Lambda_b},-s_{\Lambda_J^*})\Big),\notag\\
L_{13}&=&\frac{3}{16}\bigg((4\hat{m}_\ell^2+1)2|H^{\frac{3}{2}}_{\frac{1}{2},\frac{1}{2}}|^2+(4\hat{m}_\ell^2+3)(|H^{\frac{3}{2}}_{\frac{1}{2},-\frac{1}{2}}|^2-|H^{\frac{3}{2}}_{\frac{1}{2},\frac{3}{2}}|^2)\bigg)+\Big((s_{\Lambda_b} ,s_{\Lambda_J^*})\to (-s_{\Lambda_b},-s_{\Lambda_J^*})\Big),\notag\\
L_{21}&=&\frac{\sqrt{3}}{8}(4\hat{m}_\ell^2-1)\mathcal{R}_e(H^{\frac{3}{2}}_{\frac{1}{2},\frac{3}{2}}H^{\frac{3}{2}*}_{\frac{1}{2},-\frac{1}{2}})+\Big((s_{\Lambda_b} ,s_{\Lambda_J^*})\to (-s_{\Lambda_b},-s_{\Lambda_J^*})\Big),\notag\\
L_{22}&=&-L_{21},\notag\\
L_{31}&=&\frac{-1}{16}(4\hat{m}_\ell^2-1)\Big(3|H^{\frac{3}{2}}_{\frac{1}{2},\frac{3}{2}}|^2-10|H^{\frac{3}{2}}_{\frac{1}{2},\frac{1}{2}}|^2-16|H^{\frac{1}{2}}_{\frac{1}{2},\frac{1}{2}}|^2+5|H^{\frac{3}{2}}_{\frac{1}{2},-\frac{1}{2}}|^2+8|H^{\frac{1}{2}}_{\frac{1}{2},-\frac{1}{2}}|^2\Big)+\Big((s_{\Lambda_b} ,s_{\Lambda_J^*})\to (-s_{\Lambda_b},-s_{\Lambda_J^*})\Big),\notag\\
L_{32}&=&-(4\hat{m}_\ell^2-1)\bigg(\mathcal{R}_e(H^{\frac{3}{2}}_{-\frac{1}{2},\frac{1}{2}}H^{\frac{1}{2}*}_{-\frac{1}{2},\frac{1}{2}})-2\mathcal{R}_e(H^{\frac{3}{2}}_{\frac{1}{2},\frac{1}{2}}H^{\frac{1}{2}*}_{\frac{1}{2},\frac{1}{2}})\bigg)+\Big((s_{\Lambda_b} ,s_{\Lambda_J^*})\to (-s_{\Lambda_b},-s_{\Lambda_J^*})\Big),\notag\\
L_{33}&=&\frac{3}{16}(4\hat{m}_\ell^2-1)(|H^{\frac{3}{2}}_{\frac{1}{2},\frac{3}{2}}|^2+2|H^{\frac{3}{2}}_{\frac{1}{2},\frac{1}{2}}|^2-|H^{\frac{3}{2}}_{\frac{1}{2},-\frac{1}{2}}|^2)+\Big((s_{\Lambda_b} ,s_{\Lambda_J^*})\to (-s_{\Lambda_b},-s_{\Lambda_J^*})\Big),\notag\\
L_{41}&=&\frac{-\sqrt{2}}{2}(4\hat{m}_\ell^2-1)\bigg(\sqrt{3}\mathcal{R}_e(H^{\frac{3}{2}}_{\frac{1}{2},\frac{3}{2}}H^{\frac{1}{2}*}_{\frac{1}{2},\frac{1}{2}})+\mathcal{R}_e(H^{\frac{3}{2}}_{\frac{1}{2},-\frac{1}{2}}H^{\frac{1}{2}*}_{\frac{1}{2},\frac{1}{2}})-\mathcal{R}_e(H^{\frac{3}{2}}_{-\frac{1}{2},-\frac{1}{2}}H^{\frac{1}{2}*}_{-\frac{1}{2},\frac{1}{2}})\bigg)+\notag\\
&&\Big((s_{\Lambda_b} ,s_{\Lambda_J^*})\to (-s_{\Lambda_b},-s_{\Lambda_J^*})\Big),\notag\\
L_{42}&=&\frac{-1}{2\sqrt{2}}(4\hat{m}_\ell^2-1)\bigg(\sqrt{3}\mathcal{R}_e(H^{\frac{3}{2}}_{\frac{1}{2},\frac{3}{2}}H^{\frac{3}{2}*}_{\frac{1}{2},\frac{1}{2}})\bigg)+\Big((s_{\Lambda_b} ,s_{\Lambda_J^*})\to (-s_{\Lambda_b},-s_{\Lambda_J^*})\Big),\notag\\
L_{51}&=&-L_{21}=- L_{52},\notag\\
L_{61}&=&\frac{1}{\sqrt{2}}(4\hat{m}_\ell^2-1)\Big(\mathcal{I}_m(H^{\frac{3}{2}}_{\frac{1}{2},-\frac{1}{2}}H^{\frac{1}{2}*}_{\frac{1}{2},\frac{1}{2}})+\mathcal{I}_m(H^{\frac{3}{2}}_{\frac{1}{2},\frac{1}{2}}H^{\frac{1}{2}*}_{\frac{1}{2},-\frac{1}{2}})-\sqrt{3}\mathcal{I}_m(H^{\frac{3}{2}}_{\frac{1}{2},\frac{3}{2}}H^{\frac{1}{2}*}_{\frac{1}{2},\frac{1}{2}})\Big)-\Big((s_{\Lambda_b} ,s_{\Lambda_J^*})\to (-s_{\Lambda_b},-s_{\Lambda_J^*})\Big),\notag\\
L_{62}&=&\frac{1}{2\sqrt{2}}(4\hat{m}_\ell^2-1)\Big(\sqrt{3}\mathcal{I}_m(H^{\frac{3}{2}}_{\frac{1}{2},\frac{1}{2}}H^{\frac{3}{2}*}_{\frac{1}{2},\frac{3}{2}})-\Big((s_{\Lambda_b} ,s_{\Lambda_J^*})\to (-s_{\Lambda_b},-s_{\Lambda_J^*})\Big),\notag\\
L_{71}&=&\frac{\sqrt{3}}{8}(4\hat{m}_\ell^2-1)\bigg(\Big(\mathcal{I}_m(H^{\frac{3}{2}}_{\frac{1}{2},-\frac{1}{2}}H^{\frac{3}{2}*}_{\frac{1}{2},\frac{3}{2}})\Big)\bigg)+\Big((s_{\Lambda_b} ,s_{\Lambda_J^*})\to (-s_{\Lambda_b},-s_{\Lambda_J^*})\Big),\notag\\
L_{72}&=&L_{81}=-L_{82}=-L_{71}.\label{coe}
\end{eqnarray}

 \section{The $J/\psi$ decay process $J/\psi \to \ell^{+}\ell^{+}$}\label{jpsi}
The F*F type interaction is parametrized as 
  \begin{eqnarray}
{\cal H}_{eff}=gF^{\mu\nu} {F}^{\prime}_{\mu\nu},
 \end{eqnarray} 
 which gives  amplitude for $J/\psi\to \bar\ell \ell$ as:  
 \begin{eqnarray}
i\mathcal{M}(J/\psi\to \ell^+\ell^-)
&=&
\langle \ell^+(s_+)\ell^-(s_-)| -i gF^{\mu\nu} {F}^{\prime}_{\mu\nu}(0)|J/\psi(s_{J/\psi})\rangle\nonumber
\\
&=&\langle \ell^+(s_+)\ell^-(s_-)| -i gF^{\mu\nu} {F}^{\prime}_{\mu\nu}(0)(-i e \int d^4 x\bar \ell \gamma^\rho\ell A_\rho(x)) |J/\psi(s_{J/\psi})\rangle\nonumber\\
&=&-2eg\int \frac{d^4q}{(2\pi)^4}q^2\int d^4x e^{ix(p_{\ell^+}+p_{\ell^-}-q)}\frac{-i}{q^2}\times\bar{u}(s_-)\gamma^\mu v(s_+)\epsilon_\mu(s_{J/\psi})\notag\\
&= & 2i eg \times\bar{u}(s_-)\gamma^\mu v(s_+)\epsilon_\mu(s_{J/\psi}). 
\end{eqnarray}

The $\gamma_\mu*A^\mu$ type Hamiltonian  is given as
   \begin{eqnarray}
{\cal H}_{eff}=g_1\bar \ell \gamma^\mu\ell A^\prime_\mu .\label{ha}
 \end{eqnarray}
 The amplitude for   $J/\psi\to \bar\ell \ell$ becomes
 \begin{eqnarray}
i\mathcal{M}(J/\psi\to \ell^+\ell^-)
&=&
\langle \ell^+(s_+)\ell^-(s_-)| -i g_1\bar \ell \gamma^\mu\ell A^\prime_\mu(0)|J/\psi(s_{J/\psi})\rangle\nonumber
\\
&= & -ig_1 \times\bar{u}(s_-)\gamma^\mu v(s_+)\epsilon_\mu(s_{J/\psi}). 
\end{eqnarray}  
 Comparing the amplitudes derived by two different types of Hamiltonian, one can find a relation between the coupling constant $g$ and $g_1$ as
 \begin{eqnarray}
g=\frac{-g_1}{2 e}.
\end{eqnarray}   
 It shows that the two parametrizations are equivalent.
 \end{appendix}

\end{document}